\documentclass[a4,12pt,epsf]{article}

\usepackage{graphicx,amsmath}
\usepackage{amsmath,amsfonts,latexsym,amssymb}
\usepackage{verbatim,amsthm,graphics}
\usepackage{mathrsfs}

\def\ben{\begin{equation}}
\def\een{\end{equation}}
\def\bena{\begin{eqnarray}}
\def\eena{\end{eqnarray}}
\def\half{{1\over 2}}
\def\quater{{1 \over 4}}

\def\f(#1/#2){\frac{#1}{#2}}
\def\Frac(#1/#2){\left(\frac{#1}{#2}\right)}
\def\chris(#1-#2-#3){{\mit \Gamma}^{#1}{}_{{#2}{#3}} }
\def\tilchris(#1-#2-#3){\tilde{{\mit \Gamma}}^{#1}{}_{{#2}{#3}}}
\def\hatchris(#1-#2-#3){\hat{{\mit \Gamma}}^{#1}{}_{{#2}{#3}}}

\newcommand{\non}{\nonumber}
\theoremstyle{definition}

\newtheorem{lemma}{Lemma}

\newcommand{\dd}{{\rm d}}

\newcommand{\mr}{{\mathbb R}}
\newcommand{\mn}{{\mathbb N}}

\newcommand{\mz}{{\mathbb Z}}
\newcommand{\mq}{{\mathbb Q}}
\newcommand{\bomega}{{\underline \Omega}}
\newcommand{\btau}{{\underline \tau}}
\newcommand{\n}{{\underline m}}

\newcommand{\e}{{\rm e}}
\newcommand{\x}{{\underline x}}
\renewcommand{\pounds}{{\mathscr L}}

\begin{document}

\title{
On the `Stationary Implies Axisymmetric' Theorem for
Extremal Black Holes in Higher Dimensions}
%A Higher Dimensional Stationary Extremal Rotating Black Hole
% Must be Axisymmetric}

\author{
Stefan Hollands$^{1}$\thanks{\tt HollandsS@Cardiff.ac.uk}\: and
%%%
Akihiro Ishibashi$^{2,3}$\thanks{\tt akihiro.ishibashi@kek.jp}\:
%%%
%and Robert M. Wald$^{2}$\thanks{\tt rmwa@midway.uchicago.edu}
\\ \\
%%%
{\it ${}^{1}$School of Mathematics, Cardiff University} \\
{\it United Kingdom,} \medskip \\
%%%
{\it ${}^{2}$Cosmophysics Group, Institute of Particle and Nuclear Studies, KEK} \\
{\it Japan,} \medskip \\
%%%
{\it ${}^{3}$Perimeter Institute for Theoretical Physics} \\
{\it Canada} \\
}

\maketitle

\abstract{All known stationary black hole solutions in higher dimensions possess
additional rotational symmetries in addition to the stationary Killing field.
Also, for all known stationary solutions, the event horizon is a Killing
horizon, and the surface gravity is constant. In the case of non-degenerate
horizons (non-extremal black holes), a general theorem was previously
established [gr-qc/0605106] proving that these statements are in fact
generally true under the assumption that the spacetime is analytic,
and that the metric satisfies Einstein's equation.
Here, we extend the analysis to the case of
degenerate (extremal) black holes. It is shown that the theorem
still holds true if the vector of angular velocities
of the horizon satisfies a certain ``diophantine condition,'' which
holds except for a set of measure zero.

\bigskip
\noindent
{KEK-Cosmo-12}
}

\section{Introduction} \label{sect:1}

In a recent paper~\cite{HIW07}, we proved the following two statements
about stationary, asymptotically flat, analytic black hole solutions to
the vacuum or electrovacuum Einstein equations with a {\sl non-degenerate}
(non-extremal) event horizon for general spacetime dimension $n \ge 4$:
(i) The event horizon is in fact a Killing horizon, and (ii) if it is
rotating, then the spacetime must also be axisymmetric.
Property (i) establishes the zero-th law of black hole thermodynamics as
the surface gravity must be constant over a Killing horizon.
Property (ii) may be viewed as a ``symmetry enhancement''
theorem, as it shows that such black holes must have at least one
more symmetry than originally assumed. Statements (i) and (ii) are often
referred to as {\sl rigidity theorem}, since they imply in particular
that the horizon must be rotating rigidly relative to infinity. An alternative
proof of these statements was recently also given in~\cite{MI08}.

\medskip
The rigidity theorem was originally proved for $n=4$ dimensions
by~\cite{H72,HE,Chr97,FRW99}, and it plays a critical role in the
proof of the black hole uniqueness
theorem~\cite{Israel67,Israel68,Carter71,Robinson75,Mazur82,Bunting83}
for stationary
(electro-)vacuum black hole solutions
in $n=4$ dimensions\footnote{An alternative strategy to prove this result
bypassing the rigidity theorem was recently proposed in~\cite{IK07}. However,
this argument relies on certain restrictive extra assumptions on the geometry
besides stationarity and asymptotic flatness.}. In higher dimensions, the uniqueness
theorem no longer holds as it stands. A variety of
explicit stationary black hole solutions have been constructed
in recent years but their complete classification
is still a major open problem\footnote{
For a partial classification see~\cite{HY1,HY2}.
and also \cite{Harmark:2004rm,Harmark:2005vn}.}.
Properties (i) and (ii) therefore place an
important restriction on such black hole solutions in $n>4$.
The purpose of the present paper is to establish a version
of the rigidity theorem also for the case of {\sl degenerate}
(extremal) black holes. This case corresponds to a vanishing
Hawking temperature
and is of particular physical importance e.g. for the investigation of
the quantum properties of black holes in string theory.

\medskip
In order to explain why the proof~\cite{HIW07} does not
carry over straightforwardly to the degenerate case, let us
first recall the basic strategy of proof employed in~\cite{HIW07}.
By assumption, there is a stationary Killing vector field, $t^a$,
which is tangent to the horizon, but not null on the horizon if the
latter is rotating. The key step in the proof is to construct
another Killing field $K^a$ which is null on the horizon. This
is obtained in turn by finding a distinguished foliation of a neighborhood
of the horizon by $(n-2)$-dimensional cross sections.
To determine that special foliation, one needs to
integrate  a certain ordinary differential equation along the orbits
of the projection $s^a$ of $t^a$ onto an arbitrary
horizon cross-section, $\Sigma$. If the orbits of $s^a$ close on $\Sigma$, then
the integration of this differential equation is straightforward. In
$n=4$, the cross section $\Sigma$ is topologically a two-sphere by
the topology theorem~\cite{HE,CW94top},
implying that the orbits of
$s^a$ must necessarily close. But in higher dimensions, the orbits need not
be closed and can in fact be dense on $\Sigma$. Nevertheless, if the
horizon is non-degenerate, then a solution to the desired ordinary
differential equation can be obtained using basic methods from ergodic
theory. Unfortunately, this method of constructing the desired solution
 does not seem to generalize straightforwardly to the
case of degenerate horizons.

\medskip
In this paper, we therefore use a different
argument which is basically as follows. First, we argue that we can
decompose $s^a = \Omega_1 \psi_1^a + \dots + \Omega_N \psi_N^a$ locally on $\Sigma$, where
$N \ge 1$ and where the vector fields $\psi_i^a$ commute and have closed orbits with period $2\pi$. The
constants $\Omega_i$ can be viewed as a local definition of the angular
velocities of the horizon. We now make a Fourier decomposition
of the quantities involved in our differential equation on the
$N$-tori ${\mathbb T}^N \subset \Sigma$ generated by the $N$ vector fields $\psi_i^a$. If this is done,
then we can construct the desired solution to our differential equation
provided the vector $\underline \Omega = (\Omega_1, \dots, \Omega_N)$ satisfies
$$
|\underline \Omega \cdot \n| > |\underline \Omega| \cdot |\n|^{-q}
$$
for some number $q$ and for all but finitely many $\n \in \mz^N$. We refer to this condition
on the angular velocities as a ``diophantine condition.'' It is satisfied
for all $\underline \Omega$ except for a set of measure zero. In summary,
if the diophantine condition holds, then we can complete the proof of
statements (i) and (ii)---i.e. the rigidity theorem---in the degenerate case.
We are unsure whether this condition is a genuine restriction or an
artefact of our method of proof.

\medskip

Our paper is organized as follows. In section~\ref{sect:2}, we
prove statement (i) and (ii) in the extremal case for vacuum
black holes. In section~\ref{sect:Matterfields}, we extend these results
to include matter fields. The matter fields that we consider consist of
a multiplet of scalar fields and abelian gauge fields with a fairly general
action, including typical actions characteristic for many supergravity
theories.
As a by-product, we also generalize our previous results in the non-extremal case~\cite{HIW07}
to such theories. The rigidity theorem for theories with an additional Chern-Simons term in the action
is proved for a typical example in Appendix~\ref{sect:appe:CS}.
In section~\ref{sect:discussion} we briefly discuss further the nature of
the diophantine condition.
% The ``double null''
The decomposition of Einstein's equation used in the main part of
the paper is given in Appendices~\ref{sect:A} and~\ref{sect:B}.

\medskip
Our signature convention for $g_{ab}$ is $(-,+,+,\cdots)$.
The Riemann tensor is defined by
$R_{abc}{}^d k_d = 2\nabla_{[a} \nabla_{b]} k_c$ and
the Ricci tensor by $R_{ab} = R_{acb}{}^c$. We also set $8\pi G = 1$.

\section{Proof of the rigidity theorem in the vacuum case}
\label{sect:2}

Let $(M,g_{ab})$ be an $n$-dimensional, smooth,
asymptotically flat, stationary solution
to the vacuum Einstein equation containing a black hole.
Thus, we assume the existence in the spacetime of a
Killing field $t^a$ with complete orbits which are timelike near infinity.
Let $H$ denote the portion of
the event horizon of the black hole that lies to the
future of past null infinity ${\mathscr I}^- \cong \mr \times S^{n-2}$.
We assume that $H$ has topology $\mr \times \Sigma$, where $\Sigma$ is compact
and connected. (If $\Sigma$ is not connected, our arguments can be
applied to any connected component of $\Sigma$.) We
assume that $t^a$ is not everywhere tangent (and hence normal) to the null generators
of $H$. The event horizon $H$ is mapped into itself by a one-parameter
group of isometries generated by $t^a$.
Following our earlier paper~\cite{HIW07}, and work of
Isenberg and Moncrief~\cite{MI83,IM85}, our aim in this section is
to prove that there exists a vector field $K^a$ defined in a neighborhood
of $H$ which is normal to $H$ and on $H$ satisfies
\ben
\label{civ1}
\underbrace{\pounds_\ell \, \pounds_\ell \, \cdots \,
\pounds_\ell}_{m \,\, {\rm times}}
(\pounds_K g_{ab}) = 0, \quad m=0,1,2, \dots,
\een
where $\ell$ is an arbitrary vector field
transverse to $H$. As we shall show at the end of this section,
if we assume analyticity of $g_{ab}$ and of $H$
it follows that $K^a$ is a
Killing field.

\medskip
We shall proceed by constructing a candidate Killing field, $K^a$, and then
proving that eq.~(\ref{civ1}) holds
for $K^a$. This candidate Killing field is expected
to satisfy the following properties: (i) $K^a$ should be normal to $H$.
(ii) If we define $S^a$ by
\ben
S^a = t^a - K^a
\een
then, on $H$, $S^a$ should be tangent to cross-sections\footnote{Note that,
since $H$ is mapped into itself by the time
translation isometries, $t^a$ must be tangent to
$H$, so $S^a$ is automatically tangent to $H$. Condition (iii) requires that
there exist a foliation of $H$ by cross-sections $\Sigma(u)$ such that
each orbit of $S^a$ is contained in a single cross-section.} of $H$.
(iii) $K^a$ should commute with $t^a$.
(iv) $K^a$ should have constant surface gravity on $H$, i.e., on $H$ we should
have $K^a \nabla_a K^b = \kappa K^b$ with $\kappa$ constant on $H$,
since, by the zeroth law of black hole
mechanics, this property is known to hold
on any Killing horizon in any vacuum solution of Einstein's equation.

\medskip
We begin by choosing a cross-section $\Sigma$, of $H$.
By arguments similar to those given in the proof of proposition 4.1
of \cite{CW94}, we may assume without loss of generality that $\Sigma$
has been chosen so that each orbit of $t^a$ on $H$ intersects $\Sigma$
at precisely one point, so that $t^a$ is everywhere transverse to
$\Sigma$. We extend $\Sigma$
to a foliation, $\Sigma(u)$, of $H$ by the action of
the time translation isometries, i.e.,
we define $\Sigma(u) = \phi_u (\Sigma)$, where $\phi_u$ denotes the
one-parameter group of isometries generated by $t^a$.
Note that the function $u$ on $H$ that labels the cross-sections in this
foliation automatically satisfies
\ben
\pounds_t u = 1 \, .
\een
Next, we define $n^a$ and $s^a$ on $H$ by
\ben
t^a = n^a + s^a \, ,
\een
where $n^a$ is normal to $H$ and $s^a$ is tangent to $\Sigma(u)$.
It follows from the transversality of $t^a$ that $n^a$ is everywhere
nonvanishing and future-directed.
Note also that $\pounds_n u = 1$ on $H$.
Our strategy is to extend this
definition of $n^a$ to a neighborhood of $H$
via Gaussian null coordinates.
This construction of $n^a$ obviously
satisfies conditions (i) and (ii) above, and it also will be shown below
that it satisfies condition (iii). However,
it will, in general, fail to satisfy (iv). We shall then modify
our foliation so as to produce a new foliation $\tilde{\Sigma} (\tilde{u})$
so that (iv)
holds as well. We will then show that the corresponding
$K^a=\tilde{n}^a$ satisfies
eq.~(\ref{civ1}).

\medskip
Given our choice of $\Sigma(u)$ and the corresponding choice of $n^a$ on $H$,
we can uniquely define
a past-directed null vector field $\ell^a$ on $H$ by the requirements
that $n^a \ell_a = 1$, and that $\ell^a$ is orthogonal
to each $\Sigma(u)$.
Let $r$ denote the affine parameter on the null geodesics determined by
$\ell^a$, with $r=0$ on $H$. Let $x^A = (x^1, \dots, x^{n-2})$ be local
coordinates on an open subset of $\Sigma$. We extend these coordinates
to an open neighborhood of $H$ by demanding that they be constant along the orbits
of $n^a$ and of $\ell^a$. The coordinates $(u,r,x^A)$ that are constructed
in this manner are referred to as Gaussian null coordinates. If we cover
$\Sigma$ with an atlas of charts, then we obtain a corresponding atlas
of Gaussian null coordinates covering an open neighborhood of $H$.
The metric takes the form
\ben
\dd s^2 = 2(\dd r - r \alpha \dd u - r\beta_A \dd x^A) \dd u
        + \gamma_{AB} \dd x^A \dd x^B \, .
\label{metric:GNC}
\een
We write
\ben
\beta_a = \beta_A (\dd x^A)_a \, , \quad \gamma_{ab} = \gamma_{AB} (\dd x^A)_a (\dd x^B)_b \, ,
\een
and we note that $\beta_a, \gamma_{ab}$ are tensor fields that intrinsically defined in
a neighborhood of $H$, independent
of the choice of coordinates $x^A$ on $\Sigma$. Both these tensor fields are by
definition orthogonal to $n^a$ and $\ell^a$, meaning
$\beta_a n^a = \beta_a \ell^a = 0$ and
$\gamma_{ab} n^a = \gamma_{ab} \ell^a = 0$. It follows from the definition
of $u$  and $r$ that
\ben
\pounds_t u = 1 \, , \quad \pounds_t r = 0 \, ,
\een
and that
\ben
\pounds_t n^a = 0 \, , \quad \pounds_t \ell^a = 0 \, .
\een
It can also be shown that
\ben
\pounds_t \alpha = 0 \, , \quad
\pounds_t \beta_a = 0 \, , \quad
\pounds_t \gamma_{ab} = 0 \, .
\label{tabc}
\een
We also have
\ben
n^a = \left(\frac{\partial}{\partial u} \right)^a \,,
\quad \ell^a = \left(\frac{\partial}{\partial r}\right)^a \,\,\,,
\een
and $n^a$ and $\ell^a$ commute in particular.
Thus, we see that in Gaussian null coordinates
the spacetime metric, $g_{ab}$, is characterized by
the quantities $\alpha$, $\beta_a$, and $\gamma_{ab}$.
In terms of these quantities,
if we were to choose $K^a = n^a$, then the
condition~\eqref{civ1} will hold if and only if the conditions
\bena
\label{cid}
\underbrace{\pounds_\ell \, \pounds_\ell \, \cdots \,
\pounds_\ell}_{m \,\, {\rm times}}
\left(\pounds_n \gamma_{ab} \right) &=& 0 \, ,
\non\\
\underbrace{\pounds_\ell \, \pounds_\ell \, \cdots \,
\pounds_\ell}_{m \,\, {\rm times}}
\left(\pounds_n \alpha \right) &=& 0 \, ,
\non\\
\underbrace{\pounds_\ell \, \pounds_\ell \, \cdots \,
\pounds_\ell}_{m \,\, {\rm times}}
\left(\pounds_n \beta_a \right) &=& 0 \, ,
%\\
%\pounds_K
%\underbrace{\pounds_\ell \, \pounds_\ell \, \cdots \,
%\pounds_\ell}_{n \,\, {\rm times}}
%\left(\gamma_{AB} \dd x^A \dd x^B \right)  &=& 0 \,,
\eena
hold on $H$. The next step in the analysis is to use the Einstein
equation $R_{ab} n^a n^b = 0$ on $H$, in a manner completely in
parallel with the
4-dimensional case~\cite{HE}. This equation
is precisely the Raychaudhuri
equation for the congruence of null curves defined by
$n^a$ on $H$. It yields $\pounds_n \gamma_{ab} = 0$.
Thus,
the first equation in eq.~(\ref{cid}) holds with $m=0$.
However, $n^a$ in general fails to satisfy condition (iv) above.
Indeed, from the form, eq.~(\ref{metric:GNC}), of the metric, we see that
the surface gravity, $\kappa$, associated with $n^a$ is simply $\alpha$,
and there is no reason why $\alpha$ need be constant on $H$.
Since $\pounds_n \gamma_{ab} = 0$ on $H$,
the Einstein equation~\eqref{nAR} of Appendix~\ref{sect:A} on $H$ yields
\ben
  D_a \alpha = \frac{1}{2}  \pounds_n \beta_a \, ,
\label{dalpha}
\een
where $D_a$ denotes the derivative operator on $\Sigma(u)$, i.e.,
$D_a \alpha = {\gamma_a}^b \nabla_b \alpha$.
Thus, if $\alpha$ is not constant on $H$, then the last
equation in eq.~(\ref{cid}) fails to hold even when $m=0$.

As previously indicated, our strategy is repair this problem by choosing
a new cross-section $\tilde{\Sigma}$ so that the corresponding
$\tilde{n}^a$ arising
from the Gaussian normal coordinate construction will have constant
surface gravity on $H$. The determination of
this $\tilde{\Sigma}$ requires some intermediate constructions, to which
we now turn. First, since we already know that $\pounds_t \gamma_{ab} = 0$
everywhere and that
$\pounds_n \gamma_{ab} = 0$ on $H$, it follows immediately from the fact
that $t^a=s^a+n^a$ on $H$ that
\ben
\pounds_s \gamma_{ab} = 0 \,
\een
on $H$ (for any choice $\Sigma$). Thus, $s^a$ is a Killing
vector field for the Riemannian metric
$\gamma_{ab}$ on $\Sigma$.
Therefore the flow, $\hat \phi_\tau: \Sigma \to \Sigma$
of $s^a$ yields a one-parameter group of isometries of $\gamma_{ab}$, which coincides
with the projection of the flow $\phi_u$ of the original Killing
field $t^a$ to $\Sigma$. Furthermore, using that $\pounds_t \beta_a = 0$, it
similarly follows that
\ben\label{sbetaa}
 D_a \alpha = -\frac{1}{2} \pounds_s \beta_a \,
\een
on $H$. We next define
\ben\label{kdef}
\kappa = \frac{1}{{\rm Area}(\Sigma)} \int_\Sigma \alpha \, \dd V \, ,
\een
where $\dd V$ is the volume element on $\Sigma$ defined from $\gamma_{ab}$.
In our previous paper~\cite{HIW07}, we assumed that $\kappa \neq 0$,
i.e., that the horizon is {\sl non-degenerate}.
Here, we assume that the horizon is {\sl degenerate}, $\kappa = 0$.

\medskip
We seek a new Gaussian null coordinate system based on
a new choice $\tilde \Sigma$ of the initial cross section
such that the corresponding fields $\tilde u, \tilde r, {\tilde x}^A,
\tilde \alpha, \tilde \beta_a, \tilde \gamma_{ab}$ satisfy all the above
properties together with the additional requirement that $\tilde{\alpha} = 0$,
i.e., constancy of the surface gravity. Let us determine the conditions that
these new coordinates would have to satisfy.
Since clearly $\tilde{n}^a$ must be proportional to $n^a$, we have
\ben
\label{nfn}
\tilde n^a = f n^a \,,
\een
for some positive function $f$. Since $\pounds_t  \tilde{n}^a
= \pounds_t n^a = 0$, we must have $\pounds_t f = 0$. Since on $H$ we have
$n^a \nabla_a n^b = \alpha n^b$ and $\tilde{\alpha}$ is given by
\ben
\tilde n^a \nabla_a \tilde n^b = \tilde \alpha \tilde n^b \, .
\een
we find that $f$ must satisfy
\ben
\tilde \alpha = \pounds_n f + \alpha f = -\pounds_s f + \alpha f = 0 \, .
\label{f}
\een
The last equality provides an equation that must be satisfied by $f$ on
$\Sigma$. Writing $F = \log f$, this equation may be written alternatively
in the form
\ben
\label{Fdef}
\pounds_s F = \alpha \, .
\een
The new coordinate $\tilde{u}$ must satisfy
\ben
\label{normali}
\pounds_t \tilde u = 1 \, ,
\een
as before. However, in view of eq.~(\ref{nfn}), it also must satisfy
\ben
\pounds_n \tilde u = n^a \nabla_a \tilde{u}
= \frac{1}{f} \tilde{n}^a \nabla_a \tilde{u} = \frac{1}{f}  \, .
\een
Since $n^a = t^a - s^a$, we find that on $\Sigma$, $\tilde{u}$ must satisfy
\ben
\label{tildeu}
\pounds_s \tilde u = 1-\e^{-F}  \, .
\een
Thus, if our new Gaussian null coordinates exist, there must exist
smooth solutions to eqs.~\eqref{Fdef} and~\eqref{tildeu}, and conversely,
any solution to these equations will give us the desired new set
of Gaussian null coordinates.

\medskip
It is not difficult to show that there is always an analytic solution
$F$ to eq.~\eqref{Fdef}. To see this, we take the gradient $D_a$ of
that equation, we use that $s^a$ is a Killing field of $\gamma_{ab}$
and we use the Einstein equation~\eqref{sbetaa}. This shows that $F$ must
satisfy
\ben
\pounds_s D_a F = D_a \alpha = -\frac{1}{2} \pounds_s \beta_a \, .
\een
Taking now a divergence $D^a$ of this equation, it follows that
\ben
\pounds_s\bigg( D^a D_a F + \frac{1}{2} D^a \beta_a \bigg) = 0 \, .
\een
Thus, if we choose $F$ as a solution to the equation
$D^a D_a F = -\frac{1}{2} D^a \beta_a$, then this $F$ will
satisfy the desired equation~\eqref{Fdef}, up to
a term annihilated by $D^a D_a$, i.e. a constant,
$\pounds_s F = \alpha + {\rm const.}$ But we have
shown in our previous paper that
\ben
\lim_{T \to \infty} \frac{1}{T} \int_0^T
      \alpha \circ \hat \phi_\tau(x) \, \dd \tau = \kappa = 0 \,
\een
from which it follows that the constant vanishes. Thus, we have
constructed a solution $F$ to eq.~\eqref{Fdef}. It follows from standard
elliptic regularity results on the Laplace
operator on a compact Riemannian manifold $(\Sigma, \gamma_{ab})$
that $F$ is smooth and that it is even analytic if $\gamma_{ab}$ and
$\beta_a$ are analytic.

\medskip
We are free to add to our solution
$F$ any function $F^*$ on $H$ with the property that
$\pounds_s F^* = 0$. We take
\ben\label{Fstar}
{\rm exp} [-F^*(x)] = \lim_{T \to \infty} \frac{1}{T} \int_0^T {\rm exp}[-F \circ \hat \phi_\tau(x)] \, \dd \tau \, ,
\een
where the limit exists by the ergodic theorem~\cite{Walter}, since
$\hat \phi_\tau$ are isometries of $\gamma_{ab}$ and hence in particular
area-preserving. Again by the ergodic theorem, the right
side can also be written as the integral over the
closure of the orbit of $\hat \phi_\tau$. Using precisely the
same arguments as below in the proof of lemma~\ref{lemma:1}\footnote{
The statement follows by establishing bounds on the
derivatives of ${\rm exp}[-F^*(y)]$. These bounds are
obtained precisely as in~\eqref{partialJ},
by considering $\n = \underline{0}$ and
replacing $J(y)$ by ${\rm exp}[-F(y)]$ in that equation.}, it
is possible to show that $F^*$ is analytic. By replacing $F$ with
$F-F^*$ if necessary, we can hence achieve that our solution $F$ to
eq.~\eqref{Fdef} satisfies eq.~\eqref{Fstar}
with $\e^{-F^*}=1$. This will turn out to be convenient momentarily,
as the orbit average of the source
term in eq.~\eqref{tildeu} then vanishes.

\medskip
We now turn to eq.~\eqref{tildeu}. We note that this equation actually has
exactly the same form as eq.~\eqref{Fdef}. Also, in both cases the orbit average
of the source term on the right side vanishes. However, a difference is that, for
eq.~\eqref{tildeu}, we do not appear to have a differential relation analogous to~\eqref{sbetaa}.
Hence, it does not appear to be possible to solve that equation by the same type of
technique as eq.~\eqref{Fdef}. For this reason, we now turn to a different technique.
For this, we first consider the abelian Lie-group $\cal G$ of isometries of
$(\Sigma, \gamma_{ab})$ that is generated by the flow
$\hat \phi_\tau, \tau \in \mr$ of the vector field $s^a$.
The isometry group of any compact Riemannian manifold is known to be a
compact Lie group, so it follows that the closure $\cal K$ of $\cal G$ must
be contained in the isometry group. Being the closure of an abelian Lie-group,
$\cal K$, too, must be abelian, and hence it must be contained in a maximal
torus of the isometry group of $(\Sigma, \gamma_{ab})$. Hence, it must be
isomorphic to an $N$-torus, ${\cal K} \cong {\mathbb T}^N$, for some
$N \ge 1$. Let $\psi_1^a, \dots, \psi^a_N$, be the Killing fields on
$(\Sigma, \gamma_{ab})$
corresponding to the $N$ commuting generators of ${\mathbb T}^N$.
We assume them to be normalized so that their orbits close after $2\pi$.
Then we have
\ben\label{sadef}
s^a = \Omega^{}_1 \psi_1^a + \dots + \Omega^{}_N \psi_N^a \, ,
\een
for some numbers $(\Omega_1, \dots, \Omega_N)$, all of which
are non-zero.  If $N=1$, then the orbits of $s^a$ are closed. If
$N>1$, then the orbits of $s^a$ are not closed, and the numbers
$\Omega_i$ are linearly independent over $\mathbb Z$.
Since the choice of commuting generators of ${\mathbb T}^N$ is
arbitrary, the vector of
numbers $(\Omega_1, \dots, \Omega_N) \in \mr^N$ is unique up
to
\ben\label{ambig}
\Omega_i \to \sum_{j=1}^N A_{ij} \Omega_j \, , \quad
\pm
\left(
\begin{matrix}
A_{11} & \dots & A_{1N}\\
\vdots &       & \vdots\\
A_{N1} & \dots & A_{NN}\\
\end{matrix}
\right)
\in SL(N, \mz) \, .
\een
The Riemannian manifold $(\Sigma, \gamma_{ab})$ may
be identified with the space of null-generators of the horizon. Since this
is an invariant concept, the vector of numbers
$(\Omega_1, \dots, \Omega_N) \in \mr^N$, too, is invariantly
defined in terms of $(M,g_{ab})$, i.e., it does not depend
on our choice of $\Sigma$ up to the above ambiguity.
If it was already known that
the vector fields $\psi^a_i$ were extendible to global
Killing fields, then $\Omega_i$ would be
the corresponding angular velocities of the horizon.

\medskip
That the desired solution to eq.~\eqref{tildeu} exists is a consequence
of the following lemma:
\begin{lemma}
Let $J$ be a smooth function on $\Sigma$ with the property that
\ben\label{assumption}
0 = \lim_{T \to \infty} \frac{1}{T} \int_0^T
    J \circ \hat \phi_\tau(x) \, \dd \tau \, .
\een
Let $\bomega=(\Omega_1, \dots, \Omega_N) \in \mr^N$ [see eq.~\eqref{sadef}]
satisfy the following ``diophantine condition": There exits a  number $q$ such that\footnote{
Note that $\bomega \cdot \n \neq 0$ if $\n \neq 0$, since the entries of $\bomega$
are linearly independent over $\mathbb Z$.}
\ben
|\bomega \cdot \n| > |\bomega| \cdot |\n|^{-q}
\label{condi:dioph}
\een
holds for all but finitely many $\n \in \mz^N$.
Then the equation
\ben
\pounds_s h = J \, ,
\een
with $s^a$ as in
eq.~\eqref{sadef}, has a smooth solution $h$ on $\Sigma$.
Furthermore, if $J$ is real analytic, then the same statements
hold true and $h$ is real analytic.
\label{lemma:1}
\end{lemma}
{\em Proof:}
Let us assume that $J$ is real analytic.
It is instructive to first treat the case
$N=1$ separately. In this case, the diophantine condition is
trivially fulfilled. If $T=2\pi/ \Omega_1$, then
$\hat \phi_T(x) = x$ for all $x$ in $\Sigma$. We define
\ben
h(x) = \frac{1}{T} \int_0^T J \circ \hat \phi_\tau(x) \, \tau \dd \tau \, .
\een
This function is analytic, and we claim that it also solves the
desired differential equation. Indeed, we have
\begin{eqnarray}
\pounds_s h(x) &=& \frac{1}{T} \int_0^T
 \pounds_s J \circ \hat \phi_\tau(x) \, \tau \dd \tau \non\\
&=& \frac{1}{T} \int_0^T
    \frac{\dd}{\dd \tau} \, J \circ \hat \phi_\tau(x) \, \tau \dd \tau \non\\
&=& -\frac{1}{T} \int_0^T J \circ \hat \phi_\tau(x) \, \dd \tau
+ \frac{\tau}{T} \, J \circ \hat \phi_\tau(x) \Bigg|_{\tau=0}^{\tau = T}
\non\\
&=& J(x) \,.
\end{eqnarray}
We next treat the case $N>1$. In that case, we have
$\Omega_i/\Omega_j \notin \mq$ for
$i \neq j$, and the diophantine condition is non-trivial.
Let $\btau = (\tau_1, \dots, \tau_N) \in \mr^N/(2 \pi \mz)^N = {\mathbb T}^N$ and
let $\Phi_{\btau} \in {\rm Isom}(\Sigma)$ be the isometry of
$\Sigma$ defined as follows. For each $x \in \Sigma$ we let $\Phi_{\underline \tau}(x)$ be
the point in $\Sigma$ obtained by letting $x$ flow for parameter time $\tau_1$  along
the flow line  of the Killing field $\psi_1^a$ of $\Sigma$,
then for parameter time $\tau_2$  along
the flow line  of the Killing field $\psi_2^a$ etc. The order in
which these flows are applied does not matter as the Killing fields
mutually commute. We next define
\ben\label{fouriert}
J(x, \n) = \frac{1}{(2\pi)^N} \int_0^{2\pi} \dots \int_0^{2\pi}
                              \e^{i \n \cdot \btau}\,
                   J \circ \Phi_{\btau}(x) \, \dd \tau_1 \dots \dd \tau_N \,.
\een
The term under the integral is analytic in $(\tau_1, \dots, \tau_n)$
for each fixed $x$, so it may be analytically continued for sufficiently
small $|{\rm Im} \, \tau_i| < c_i(x)$, where $c_i(x)$ is positive.
Because $\Sigma$ is compact, it follows that the infimum $c_i$ of
$c_i(x)$ as $x$ ranges over $\Sigma$ and as $i$ ranges from $1, \dots, N$ is
a positive constant. By shifting the contours of integration to
${\rm Im} \, \tau_i = {\rm sign}(m_i) c_i$, it then follows that
\ben\label{fouriert1}
J(x, \n) = \frac{1}{(2\pi)^N} \int_{\pm ic_1}^{2\pi \pm ic_1} \dots
 \int_{\pm ic_N}^{2\pi \pm i c_N} \e^{i \n \cdot \btau} \,
 J \circ \Phi_{\btau}(x) \, \dd \tau_1 \dots \dd \tau_N \,,
\een
and therefore that (setting $c = \sqrt{N} \, {\rm inf}\{c_i \, ; \,\, i=1,\dots,N\}$)
\bena
|J(x, \n)| &\le&
\e^{-c|\n|} \, \sup \{ |J \circ \Phi_{\btau}(x)| \, ; \quad x \in \Sigma \, ,
\,\,  0 \le {\rm Re} \, \tau_i \le 2\pi \, ,
\,\, \,\, |{\rm Im} \, \tau_i| = c_i \} \non\\
&=& {\rm const.} \, \e^{-c|\n|}\, ,
\eena
for all $\n \in \mz^N$, uniformly in $x$. We now set
\ben\label{hxdef}
h(x) = i\sum_{\n \in \mz^N \setminus \underline 0} \frac{J(x,\n)}{\bomega \cdot \n} \, .
\een
We claim that this is the desired solution. Let us first check that this
is well-defined for all $x$. In view of eq.~\eqref{condi:dioph}, we can
estimate $|h(x)|$ by pulling the absolute values inside
the series~\eqref{hxdef}, to obtain
\bena
|h(x)| &\le& \sum_{\n \in \mz^N \setminus \underline 0} \frac{{\rm const.} \, \e^{-c|\n|}}{\bomega \cdot \n}
\le \frac{{\rm const.}}{|\bomega|} \sum_{\n \in \mz^N \setminus \underline 0} |\n|^q \, \e^{-c|\n|} \non\\
&\le& \frac{{\rm const.} \, q!}{ c^n \, |\bomega|} \, .
\eena
This estimate is uniform in $x \in \Sigma$.
Hence, the series~\eqref{hxdef} for
$h(x)$ converges absolutely, uniformly in $x$. We would next like
to show that $h(x)$ is real analytic. For this, we recall that if
a function $\psi$ on $\mr^k$ is real analytic near the origin in $\mr^k$,
then there is an $r>0$ and a $K>0$ such that
\ben\label{analy}
|\partial_\alpha \psi(y)| \le K^{|\alpha|} \alpha! \,\,\, ,
\een
for all $y$ in an open ball of radius $r$ around the origin. Here we use
the multi-index notation $\alpha = (\alpha_1, \dots, \alpha_k) \in
{\mathbb N}_0^k$,
\ben
\partial^\alpha = \frac{\partial^{|\alpha|}}{(\partial y^1)^{\alpha_1} \cdots
(\partial y^k)^{\alpha_k}} \, , \quad |\alpha| = \sum_i \alpha_i \, ,
\quad \alpha! = \prod_i \alpha_i! \,\,\,\, .
\een
This statement follows from the multi-dimensional generalization of
the Cauchy integral representation of an analytic function. Conversely,
if eq.~\eqref{analy} holds, then $\psi$ is analytic near the origin.
Now let $\psi$ be a real analytic function on
$\Sigma$, choose a point $x_0 \in \Sigma$,
and let $y^1, \dots, y^{n-2}$ be Riemannian normal coordinates
centered at $x_0$. Then there exist $K,r>0$ such that
eq.~\eqref{analy} holds for $\psi(y)$ for all $y$ in a
ball of radius $r$ around the origin (here we identify a neighborhood
of $x_0$ with an open neighborhood of the origin of the Riemann
normal coordinates). Furthermore, since $\Sigma$ is compact,
we may choose $K,r$ to depend only on $\psi$, but not on
the choice of $x_0$. If $c_i>0$ are as above and $\underline{c} = ({\rm sign}(m_1)c_1, \dots, {\rm sign}(m_N) c_N)
\in \mr^N$, we have
\ben
\partial^\alpha (J \circ \Phi_{\underline{\tau}+i\underline{c}}(y))
= \partial^\alpha(J \circ \Phi_{i\underline{c}} \circ
\Phi_{\underline{\tau}}(y)) =
(\partial^{\prime \,\alpha} \psi)(y')  \, ,
\een
where the derivatives in the last expression are taken with respect
to the Riemann normal coordinates centered at the image
of $x_0$ under the isometry $\Phi_{\underline{\tau}}$,
and where $y'$ is the image of $y$, identified with the corresponding
Riemann normal coordinates. In the last step,
we have used that, because $\Phi_{\underline{\tau}}$ is an isometry,
it takes Riemann normal coordinates to Riemann normal coordinates.
Furthermore, we have defined the real analytic function $\psi$
on $\Sigma$ by $\psi = J \circ  \Phi_{i\underline{c}}$. We now
apply the above estimate~\eqref{analy} to obtain
\ben
\Big| \partial^\alpha(J \circ \Phi_{\underline{\tau}+i\underline{c}}(y)) \Big|
\le K^{|\alpha|} \alpha! \, ,
\een
for all $y$ in a ball of radius $r$.
As above, we next shift the contour of the $\underline{\tau}$ integration
in the expression for $\partial_\alpha J(y, \n)$ by $i\underline{c}$,
to arrive at
\bena\label{partialJ}
|\partial^\alpha J(y, \n)| &=& \frac{\e^{-\underline{c} \cdot \n}}{(2\pi)^N}
\bigg|
\int_0^{2\pi} \dots \int_0^{2\pi} \e^{i \n \cdot \btau}
\,\, \partial^\alpha (J \circ \Phi_{\btau + i\underline{c}}(y)) \,
\dd \tau_1 \dots \dd \tau_N
\bigg| \non \\
&\le& \e^{-c|\n|} K^{|\alpha|} \alpha! \,\,\,\, .
\eena
Substituting this bound into the series for $\partial_\alpha h(y)$
and bounding each term in this series by its absolute value,
we obtain $|\partial^\alpha h(y)| \le C^{|\alpha|} \alpha!$ for
some constant $C>0$ and all $y$ in a ball of radius $r$.
Hence, $h(y)$ is analytic, as we desired to show.

\medskip
We finally need to check that $h(x)$ as defined above satisfies the desired differential
equation. For this, we first note that $J(x, \underline 0) = 0$. Indeed,
since $\Omega_i/\Omega_j \notin \mathbb Q$, we know that the orbit of
\ben
\mr \to {\mathbb T}^N, \quad t \mapsto (t\Omega_1, \dots, t\Omega_N) \, \mod \, (2\pi \mz)^N
\een
is dense in ${\mathbb T}^N$, so application of the ergodic theorem
(see e.g.~\cite{Walter}) gives
\bena
J(x, \underline 0) &=& \frac{1}{(2\pi)^N} \int_0^{2\pi} \dots \int_0^{2\pi}
\, J \circ \Phi_{\btau}(x) \, \dd \tau_1 \dots \dd \tau_N \non\\
&=& \lim_{T \to \infty} \frac{1}{T} \int^T_0
   J \circ \Phi_{(t\Omega_1, \dots, t\Omega_N)}(x) \, \dd t  \,.
\eena
On the other hand, $\Phi_{(t\Omega_1, \dots, t\Omega_N)}(x)$ is
by definition equal to $\hat \phi_t(x)$. Hence,
in view of our assumption~\eqref{assumption},
we have $J(x, \underline 0) = 0$. Next, we calculate
\bena
\pounds_s J(x, \n) &=& \frac{1}{(2\pi)^N} \int_0^{2\pi} \dots \int_0^{2\pi} \e^{i \n \cdot \btau}
\, \pounds_s J \circ \Phi_{\btau}(x) \, \dd \tau_1 \dots \dd \tau_N \nonumber\\
&=& \frac{1}{(2\pi)^N} \int_0^{2\pi} \dots \int_0^{2\pi} \e^{i \n \cdot \btau}
\, \bigg(\Omega_1 \frac{\partial}{\partial \tau_1} + \dots + \Omega_N \frac{\partial}{\partial \tau_N}
 \bigg) J \circ \Phi_{\btau}(x) \, \dd \tau_1 \dots \dd \tau_N \nonumber\\
&=& -i \, \n \cdot \bomega \, J(x, \n) \, .
\eena
Using $J(x, \underline 0)=0$, we then have
\ben
\pounds_s h(x) = i \sum_{\n \in \mz^N \setminus \underline 0} \frac{\pounds_s J(x, \n)}{\bomega \cdot \n} =
\sum_{\n \in \mz^N} J(x, \n) = J \circ \Phi_{\btau = \underline 0}(x) = J(x) \, .
\een
This concludes our proof in the case when $J$ is real analytic.

\medskip
Next, suppose $J$ is only smooth. Then the argument in the case $N=1$ is unchanged and
gives a smooth solution $h$. In the diophantine case $N>1$, we now have for any $k,l \in \mn_0$ an estimate
\ben
|\Delta^l \, J(x, \n)| \le {\rm const.} \, (1+|\n|)^{-k}
\een
for a constant only depending on $k,l$, where $\Delta = D^a D_a$. It follows again from
the diophantine condition that
the sum~\eqref{hxdef} for $h(x)$ and the corresponding sums for $\Delta^l \, h(x)$ converge
uniformly for all $l$. Thus, $|\Delta^l \, h(x)|$ is uniformly bounded and hence
$h$ is in any of the Sobolev spaces $W^{p,l}(\Sigma, \dd V)$,
and therefore smooth. That $h(x)$ satisfies the desired differential equation
follows as in the analytic case.
\qed

\bigskip
The lemma shows that the desired new Gaussian null coordinates
$(\tilde u, \tilde r, \tilde x^A)$ and corresponding foliation $\tilde \Sigma(
\tilde r, \tilde u)$ exist under the assumptions stated there.
For the rest of the paper, we assume that these hold.
Now let $K^a = \tilde{n}^a$. We have previously shown that
$\pounds_{\tilde{n}} \tilde \gamma_{ab}=0$ on $H$, since this relation holds
for any choice of Gaussian null coordinates. However, since our new
coordinates have the property that
$\tilde \alpha = 0$ on $H$, we clearly have that
$\pounds_{\tilde{n}} \tilde \alpha = 0$ on $H$.
Furthermore, for our new coordinates,
eq.~(\ref{dalpha}) immediately yields
$\pounds_{\tilde{n}} \tilde \beta_a = 0$ on $H$.
Thus, we have proven that all of the relations in eq.~(\ref{cid}) hold
for $m=0$.

\medskip
We next prove that the
equation $\pounds_{\tilde \ell} \, \pounds_{\tilde n}  \tilde
\gamma_{ab} = 0$ holds on $H$. Using what
we already know about
$\tilde \beta_a, \tilde \gamma_{ab}$
and taking the Lie-derivative $\pounds_{\tilde n}$ of the components
of the Einstein equation tangent to $\tilde \Sigma(
\tilde r, \tilde u)$
(see eq.~(\ref{ABR}) of Appendix~\ref{sect:A}), we get
\ben
0 = \pounds_{\tilde n}
    \pounds_{\tilde n} \pounds_{\tilde \ell}
           \tilde \gamma_{ab}
          %+ {\kappa} \, \pounds_{\tilde \ell}\tilde \gamma_{ab}
     \,,
\label{eq:nlgamma}
\een
on $H$.
Since $t^a = \tilde n^a + \tilde s^a$, with $\tilde s^a$ tangent
to $\tilde \Sigma(\tilde{u})$, and since all quantities appearing
in eq.~(\ref{eq:nlgamma}) are Lie derived by $t^a$,
we may replace in this equation
all Lie derivatives $\pounds_{\tilde n}$ by $- \pounds_{\tilde s}$.
Hence, we obtain
\ben
\label{31}
0 = \pounds_{\tilde s} \pounds_{\tilde s} \pounds_{\tilde \ell}
  \tilde \gamma_{ab}
 %- {\kappa} \pounds_{\tilde \ell} \tilde \gamma_{ab}
 %\right]
 \,,
\een
on $\tilde \Sigma$.
Now, write $L_{ab} = \pounds_{\tilde \ell} \tilde \gamma_{ab}$. We
fix $x_0 \in \tilde{\Sigma}$ and view eq.~(\ref{31}) as an
equation holding at $x_0$ for the pullback, $\hat{\phi}^*_\tau L_{ab}$,
of $L_{ab}$ to $x_0$, where
$\hat{\phi}_\tau: \tilde \Sigma \to \tilde \Sigma$ now
denotes the flow of $\tilde s^a$. Then eq.~(\ref{31}) can be rewritten as
\ben
\frac{\dd^2}{\dd \tau^2} \, \hat{\phi}^*_\tau L_{ab} = 0 \,.
\een
Integration of this equation yields
\ben
\label{i}
\frac{1}{\tau} ( {\hat{\phi}}_\tau^* \, L_{ab}
- L_{ab} )
= C_{ab} \,,
\een
where $C_{ab}$ is a tensor at $x_0$ that is independent of $\tau$.
However, since $\hat{\phi}_\tau$ is an isometry, each orthonormal frame
component of ${\hat{\phi}}_\tau^* \, L_{ab}$ at $x_0$ is uniformly bounded
in $\tau$ by $\sup\{(L^{ab} L_{ab}(x))^{1/2}; \,\, x \in \tilde \Sigma\}$. Consequently,
the limit of eq.~(\ref{i}) as $\tau \rightarrow \infty$ immediately yields
\ben
C_{ab} = 0 \, .
\een
Thus, we have
$\pounds_{\tilde s} \pounds_{\tilde \ell} \tilde \gamma_{ab}= 0$,
and therefore
$\pounds_{\tilde n} \pounds_{\tilde \ell} \tilde \gamma_{ab}
=\pounds_{\tilde \ell} \, \pounds_{\tilde n}  \tilde
\gamma_{ab} = 0$ on $H$,
as we desired to show.

\medskip
Thus, we now have shown that the first equation in~\eqref{cid}
holds for $m=0,1$, and that the other equations hold for $m=0$,
for the tensor fields associated with the ``tilde"
Gaussian null coordinate system, and $K^a = \tilde n^a$.
In order to prove that eq.~(\ref{cid}) holds for all $m$, we proceed
inductively. Let $M \ge 1$, and assume inductively that
the first of equations~\eqref{cid} holds for all $m \le M$, and that the
remaining equations hold for all $m \le M-1$. Our task is
to prove that these statements then also hold when $M$ is
replaced by $M+1$.
To show this, we apply the operator $(\pounds_{\tilde \ell})^{M-1}
\pounds_{\tilde n}$
to the Einstein equation $R_{ab} \tilde n^a
\tilde \ell^b=0$ (see eq.~(\ref{nlR})) and restrict to $H$.
Using the inductive hypothesis,
one sees that
$(\pounds_{\tilde \ell})^M (\pounds_{\tilde n} \tilde \alpha)=0$ on $H$,
thus establishes the second equation in~\eqref{cid} for $m \le M$.
%the quantity $(\pounds_{\tilde \ell})^M (\pounds_{\tilde n} \tilde \alpha)$
%can be written in terms of quantities that are Lie derived
%by $\tilde n$ on $H$. This establishes the third equation
%in~\eqref{cid} for $m \le M$.
%
Next, we apply the operator $(\pounds_{\tilde \ell})^{M-1}
\pounds_{\tilde n}$ to the components of Einstein's equation $R_{ab}
\tilde \ell^b=0$ tangent to $\tilde \Sigma(
\tilde r, \tilde u)$ (see eq.~(\ref{lAR})),
and restrict to $H$. Using the inductive
hypothesis, one sees that
$(\pounds_{\tilde \ell})^M (\pounds_{\tilde n} \tilde \beta_a)=0$
on $H$, thus establishes the third equation in~\eqref{cid} for $m \le M$.
%
%the quantity $(\pounds_{\tilde \ell})^M (\pounds_{\tilde n} \tilde \beta_a)$
%can be written in terms of quantities that are Lie
%derived by $\tilde n^a$ on $H$. This establishes the second equation
%in~\eqref{cid} for $m \le M$.
%
Next, we apply the operator $(\pounds_{\tilde \ell})^M \pounds_{\tilde n}$
to the components of Einstein equation tangent to $\tilde \Sigma(
\tilde r, \tilde u)$
%$R_{ab}
%(\partial/\partial \tilde x^A)^a(\partial/\partial \tilde x^B)^b
%=0$
(see eq.~(\ref{ABR})), and restrict to $H$. Using the inductive
hypothesis and the above results
$(\pounds_{\tilde \ell})^M (\pounds_{\tilde n} \tilde \alpha)=0$
and $(\pounds_{\tilde \ell})^M (\pounds_{\tilde n} \tilde \beta_a)=0$,
one sees that the tensor field
$S_{ab} \equiv (\pounds_{\tilde \ell})^{M+1} \tilde \gamma_{ab}$
satisfies a differential equation of the form
\ben
\pounds_{\tilde n} \pounds_{\tilde n} S_{ab}
                   = 0
\een
on $H$. By the same argument as given above for $L_{ab}$,
it follows that $\pounds_{\tilde n} \, S_{ab} = 0$.
This establishes the first equation
in~\eqref{cid} for $m \le M+1$, and closes the induction loop.

\medskip
Thus, we have shown~\eqref{civ1} for our choice of $K^a$.
In the analytic case,
since $g_{ab}$ and $K^a$ are analytic, so is $\pounds_K g_{ab}$.
It follows immediately from the fact that this quantity and all of its
derivatives vanish at any point of $H$ that $\pounds_K g_{ab} = 0$ where
defined, i.e., within the region where the Gaussian null coordinates
$(\tilde{u}, \tilde{r}, \tilde{x}^A)$ are defined. This proves existence
of a Killing field $K^a$ in a neighborhood of the horizon. We
may then extend $K^a$ by analytic continuation.
Now, analytic continuation
need not, in general, give rise to a single-valued extension,
so we cannot conclude that there exists a Killing field
on the entire spacetime. However,
by a theorem of Nomizu~\cite{Nomizu60} (see also~\cite{Chr97}),
if the underlying domain is simply connected,
then analytic continuation does give rise to a single-valued extension.
By the topological censorship theorem~\cite{Galloway99,Galloway01},
the domain of outer
communication has this property. Consequently, there exists
a unique, single valued extension of $K^a$ to the domain of outer
communication, i.e., the exterior of the black hole
(with respect to a given end of infinity). Thus, in the analytic case,
we have proven the following theorem:

\paragraph{Theorem 1:} Let $(M,g_{ab})$
be an analytic,
asymptotically flat $n$-dimensional solution of the vacuum Einstein equations
containing a black hole and possessing a
Killing field $t^a$ with complete orbits which are timelike near infinity.
Assume that the event horizon, $H$, of the black hole is analytic and is
topologically $\mr \times \Sigma$, with $\Sigma$ compact and connected,
and that $\kappa = 0$ (where $\kappa$ is defined by eq.~\eqref{kdef} above).
Let $\underline \Omega = (\Omega_1, \dots, \Omega_N)$ be the angular
velocities associated with projection of $\phi_\tau$ onto $\Sigma$,
see eq.~\eqref{sadef}. If these
satisfy the diophantine condition
\ben\label{dioph}
|\bomega \cdot \n| > |\bomega| \cdot |\n|^{-q}
\een
for some number $q$ and for all but finitely many $\n \in \mz^N$, then there
exists a Killing field $K^a$ whose orbits are tangent to the null-generators
of $H$.

\medskip
\noindent
{\bf Remarks:} \,
{\bf (1)} Note that the diophantine condition is trivially satisfied when
$N=1$, i.e., when the one-parameter group of symmetries $\phi_\tau$ associated
with $t^a$ maps the horizon generators to themselves after some fixed
period $T$. For $N>1$, the diophantine condition is non-trivial.
We will discuss it in some more detail in section~\ref{sect:discussion}.

\noindent
{\bf (2)} If the diophantine condition is satsified for $\underline \Omega$,
then it is also satisfied for
$A \underline \Omega$ when $\pm A \in SL(N,\mz)$. Thus,
the diophantine condition is invariant under changes of the
form~\eqref{ambig}, which as we discussed, constitute the only ambiguity
in our definition of $\underline \Omega$ for the given spacetime.

\medskip
If we are in the situation described in Theorem~1, we can apply the same type
of reasoning as in our previous paper~\cite{HIW07} to extend
the rotational Killing fields $\tilde \psi^a_i$ in the decomposition
$\tilde s^a = \Omega_1^{} \tilde \psi^a_1 + \dots + \Omega_N^{}
\tilde \psi_N^a$ [see eq.~\eqref{sadef}] to Killing fields on the entire
exterior of the spacetime, i.e., we have the following theorem.

\medskip
\paragraph{\bf Theorem 2:} Let $(M,g_{ab})$ be an analytic, asymptotically
flat $n$-dimensional solution of the vacuum Einstein equations containing
a black hole and possessing a Killing field $t^a$ with complete
orbits which are timelike near infinity. Assume that the event horizon, $H$,
of the black hole is analytic and is topologically $\mr \times \Sigma$,
with $\Sigma$ compact and connected, and that $\kappa = 0$.
As above, assume that $(\Omega_1, \dots, \Omega_N)$ [see eq.~\eqref{sadef}]
satisfy the diophantine condition~\eqref{dioph}.
If $t^a$ is not tangent to the generators of $H$, then there exist
mutually commuting Killing fields
$\tilde \psi^a_{1}, \dots, \tilde \psi^a_{N}$ (where $N\ge 1$) with
closed orbits with period $2 \pi$ which are defined
in a region that covers $H$ and the entire domain of outer communication.
Each of these Killing fields commutes with $t^a$, and $t^a$ can be written as
\ben
  t^a = K^a + \Omega_1^{} \tilde \psi_{1}^a + \dots
            +  \Omega_N^{} \tilde \psi_{N}^a \, ,
\een
where $K^a$ is the horizon Killing field whose existence is guaranteed by
Theorem~1.

\medskip
\noindent
{\bf Remarks:} \, {\bf (1)} If the spacetime is asymptotically flat
in the standard sense with asymptotic infinity of type $S^{n-2}$,
then there can be at most $N=[(n+1)/2]$ (we mean the integer part of a number)
mutually commuting Killing fields including the stationary Killing
field. For example, Myers-Perry black holes~\cite{MP86} in arbitrary $n>4$
possess a stationary Killing field plus $[(n-1)/2]$ rotational
Killing symmetries with angular velocities $\Omega_{i}$,
$i= 1,\dots , [(n-1)/2]$. These solutions admit a regular
extremal (degenerate horizon) limit for a wide range of the parameters
of the solutions, for example when all the angular velocities are equally large.
However, note that when a Myers-Perry hole has only a single non-vanishing
angular momentum, the horizon becomes singular in the extremal limit for $n=5$,
and for $n \geq 6$, there is no extremal limit; the angular velocity can
be arbitrary large in that case.
A black ring solution~\cite{ER02,Pomeransky:2006bd} in $n=5$ which possesses 3 mutally
commuting Killing fields also admits a regular extremal limit if it has
two non-vanishing angular velocities. For more details on higher dimensional,
extremal black holes see e.g.
\cite{Kunduri:2007vf,ElvangRodriguez08,Figueras:2008qh,Kunduri:2008rs,ER08},
and references therein.

{\bf (2)} If the black hole is non-rotating, i.e. if $t^a$ is tangent to
the null generators of $H$, then the solution is static~\cite{SW92}.
The same result also holds for Einstein-Maxwell theory~\cite{SW92}, and
more generally presumably also for many of the Einstein-Matter theories
described in the next section. In the non-extremal case, the uniqueness theorems~\cite{Gibbons02a,Gibbons02b} for static
Einstein-Maxwell-Dilaton black hole solutions then apply. In the extremal case
uniqueness of higher dimensional, static Einstein-Maxwell black hole
solutions was shown in~\cite{Rogatko:2003kj}.

\section{Matter fields}  \label{sect:Matterfields}
So far we have focused on vacuum solutions to the Einstein equations
for the sake of simplicity. In this section we generalize our
results to include certain types of matter fields. We consider
theories containing scalar fields $\phi$ taking values in
a target space manifold $X$ with positive definite metric $f_{ij}(\phi)$
and vector fields $A_a$ taking values in a vector bundle
over $X$ with positive definite vector bundle metric $h_{IJ}(\phi)$.
We write the components of the scalar and vector fields
as $\phi^i$ and $A^I_a$ respectively. We take the action to be
\bena
 S = \int \dd^n x \sqrt{- g}
\Big(
  R -\half f_{ij}(\phi) g^{ab} \nabla_{a} \phi^i \nabla_b \phi^j
    - U(\phi)
    - \frac{1}{4} h_{IJ}(\phi)
      g^{ac} g^{bd} F^I_{ab} F^J_{cd}
\Big) + S_{\rm top}\,,
\label{action}
\eena
where $F^I_{ab} = \nabla_a A^I_b - \nabla_b A^J_a $, where $U$ is a
potential, and where $S_{\rm top}$ denotes a topological term. A
typical example for such a term is a Chern-Simons action.
It does not affect the form of stress-energy
tensor but it can modify the equation of motion for the gauge field,
eq.~(\ref{eom:gauge}). In this section we will discard the
topological term for simplicity. But we will discuss the minimal
supergravity in $n=5$ dimensions as an example of a theory with a
Chern-Simons term in appendix~\ref{sect:appe:CS}.

\medskip
The above class of theories obviously includes the case of pure gravity with
a cosmological constant, which corresponds to
solutions with constant $\phi$. It also includes
many interesting supergravity theories in $5$-(and $4$)-dimensions
arising from supergravity theories in $11$-dimensions and string theories
in $10$-dimensions by appropriate dimensional reductions.
In the latter case, one must include a topological term.
%[AKIHIRO: I DO NOT UNDERSTAND THE FOLLOWING REMARK
%Note, however, that for supersymmetric theories,
%one can construct a Killing vector field which is non-spacelike
%everywhere outside the black hole and hence, in particular,
%must be {\sl normal} to the event horizon.
%(In that case we are already done.) Therefore, in principle, we are
%concerned with {\sl non-supersymmetric} theories described by
%the above action.]

\medskip
Varying the action eq.~(\ref{action}) gives the following
equations of motion:
\bena
% \bullet
&& R_{ab}
%  = T_{ab}-\frac{T}{n-2}g_{ab}
% \non
% \\
% && \quad
 = f_{ij} (\phi) \nabla_a \phi^i \nabla_b \phi^j
   + h_{IJ}(\phi) g^{cd} F^I_{ac}F^J_{bd}
   + \frac{2}{n-2}g_{ab}
     \Bigg[
        U(\phi) - \frac{1}{4}h_{IJ}(\phi)F^I_{cd}F^{J cd}
     \Bigg] \,,
\label{eom:grav}
\\
% \bullet
&&
   \nabla_a (f_{ij} (\phi) \nabla^a \phi^j)
 - \half f_{jk |i} \nabla^a \phi^j \nabla_a \phi^k
 - U_{| i}
 - \frac{1}{4} h_{IJ|i}F^I_{ab}F^{J ab} = 0 \,,
\label{eomq:scalar}
\\
% \bullet
&&
 \nabla_c\Bigg[ h_{IJ}(\phi) F^{J ca} \Bigg] =0 \,,
\label{eom:gauge}
% \\
% \bullet
% &&
%  \nabla_{[a} F^{J}_{bc]} =0  \,,
% \label{eq:Bianchi}
\eena
and the Bianchi identities,
\ben
 \nabla_{[a} F^{J}_{bc]} =0  \,,
\label{id:Bianchi}
\een
where here and in the following the {\sl vertical stroke} denotes
the derivative with respect to a scalar field component, $\phi^i$, e.g.,
$f_{jk |i} = \partial f_{jk}(\phi)/ \partial \phi^i$.

\medskip
We now consider a stationary black hole solution in the above theory
with corresponding Killing field $t^a$, that is $\pounds_t g_{ab} = 0$.
We also assume that the other fields are invariant under $t^a$, that is
$\pounds_t \phi^i =0 \,,\; \pounds_t F^J_{ab} = 0$, and that all fields
are real analytic. Which asymptotic conditions on the dynamical fields are
reasonable in the above theory will in general depend on the precise choice
of the potential $U(\phi)$ and the metrics $f_{ij}(\phi)$, $h_{IJ}(\phi)$.
In the vacuum case, we assumed asymptotic flatness for the metric with
standard infinity ${\mathscr I}^\pm \cong S^{n-2} \times \mr$.
This assumption was used implicitly to show that $t^a$ does not vanish
on $H$, a fact which we needed to obtain the desired foliation $\Sigma(u,r)$
in our construction of the Gaussian null coordinates. Asymptotic flatness was
also implicitly used in the proof of Theorem~2, in combination with
the topological censorship theorem~\cite{Galloway99}.
Here, it was needed in order to establish that the exterior of the black hole
is a simply connected manifold, which in
turn is essential in order to be able to analytically extend the Killing fields $K^a$ and
$\psi_i^a$ to the full exterior of the black hole in a single valued way, cf.~\cite{HIW07} for the
details of this argument. In the present section, we will
simply assume that $t^a$ is nowhere vanishing on $H$,
and that the exterior is simply connected. As in the vacuum case, we also
assume that the black hole is rotation, i.e. that $t^a$ is not everywhere
tangent to the null generators of $H$. For the case when the orbits of $t^a$ are
tangent to the generators see Remark~2 following Theorem~2.

\medskip
As in the vacuum case, we distinguish between extremal and non-extremal black
holes. In the {\sl non-extremal case} we will show that, if the orbits of
$t^a$ are not everywhere tangent to the null generators
of the horizon $H$, then the analogues of Theorems~1 and~2
hold without any restrictions on the vector of angular velocities
$\underline \Omega$. This generalizes previous results in~\cite{HIW07} to
the above type of theories.
In the {\sl extremal case} we will show the same result under the additional
assumption that the vector of angular velocities $\underline \Omega$ verifies
the diophantine condition given in the statement of Theorem~1. 

\medskip
Let us now explain how the desired additional Killing field $K^a$ described
in Theorems~1 and~2 is constructed in the above types of theories.
By analogy to the vacuum case, we must now show that
\bena
 \underbrace{\pounds_\ell \, \pounds_\ell \, \cdots \,
             \pounds_\ell
            }_{m \,\, {\rm times}}
   \left(\pounds_K g_{ab} \right) = 0 \,, \quad
 \underbrace{\pounds_\ell \, \pounds_\ell \, \cdots \,
             \pounds_\ell
            }_{m \,\, {\rm times}}
   \left(\pounds_K \phi^i \right) = 0 \,, \quad
 \underbrace{\pounds_\ell \, \pounds_\ell \, \cdots \,
             \pounds_\ell
            }_{m \,\, {\rm times}}
   \left(\pounds_K F^I_{ab} \right) = 0 \,.
\eena
Again, we first introduce a Gaussian null
coordinate system $(u,r,x^A)$ adapted to the horizon geometry, and
we seek to adjust the remaining freedom in choosing this coordinate system
in such a way that the desired $K^a$ is given by $n^a = (\partial/\partial u)^a$.

\medskip
To do this, it is convenient to first decompose the components of $F^I_{ab}$
with respect to the Gaussian null coordinate system. For
this, we define
\ben
 F^I_{ab} \: n^a \ell^b = S^I \,, \quad
 F^I_{ac} \: n^a p^c{}_b = V^I_b \,, \quad
 F^I_{ac} \: \ell^a p^c{}_b = W^I_b \,, \quad
 F^I_{cd} \: p^c{}_a p^d{}_b  = U^I_{ab} \,,
\een
where $p^a{}_b$ projects on the surfaces $\Sigma(u,r)$ of
constant $u,r$, cf. Appendix~\ref{sect:A} for details.
The field equations are written in terms of these
variables and $\gamma_{ab}, \beta_a, \alpha$ in
Appendix~\ref{sect:B}. It immediately follows from
$\pounds_t F^J_{ab} = 0$ that
$
\pounds_t S^I= 0 \,, \:
\pounds_t V^I_a= 0 \,, \:
\pounds_t W^I_a= 0 \,, \:
\pounds_t U^I_{ab}= 0
$.
Our task is now to show eqs.~(\ref{cid}) and
\bena
\underbrace{\pounds_\ell \, \pounds_\ell \, \cdots \,
\pounds_\ell}_{m \,\, {\rm times}}
\left(\pounds_n \phi^i  \right) &=& 0 \,,
\label{cid:s}
\\
\label{cid4}
\underbrace{\pounds_\ell \, \pounds_\ell \, \cdots \,
\pounds_\ell}_{m \,\, {\rm times}}
\left(\pounds_n S^I  \right) &=& 0 \, ,
\non\\
\underbrace{\pounds_\ell \, \pounds_\ell \, \cdots \,
\pounds_\ell}_{m \,\, {\rm times}}
\left(\pounds_n V^I_a \right) &=& 0 \, ,
\non\\
\underbrace{\pounds_\ell \, \pounds_\ell \, \cdots \,
\pounds_\ell}_{m \,\, {\rm times}}
\left(\pounds_n W^I_a \right) &=& 0 \, ,
\non\\
\underbrace{\pounds_\ell \, \pounds_\ell \, \cdots \,
\pounds_\ell}_{m \,\, {\rm times}}
\left(\pounds_n U^I_{ab} \right) &=& 0 \,,
\eena
for a suitable choice of our Gaussian null coordinate system.

\medskip
First, we consider the Raychaudhuri equation for a congruence of null geodesic
generators of the even horizon $H$, i.e. the Einstein equations contracted with $n^a n^b$:
\ben
 \frac{\dd }{\dd \lambda} \theta
 = - \frac{1}{n-2}\theta^2 - \widehat{\sigma}^{ab} \widehat{\sigma}_{ab}
   - f_{ij} (\pounds_n\phi^i) \pounds_n \phi^j
   - h_{IJ}q^{ab}V^I_aV^J_b \,,
\een
where $\lambda$ is an affine parameter of null geodesic generators of $H$
and where $\theta$ and $\widehat{\sigma}_{ab}$ denote,
respectively, the expansion and the shear of the null geodesic generators.
Because the terms on the right-hand side are negative
definite\footnote{Here it enters that the target space metrics
$h_{IJ}$ and $f_{ij}$ are positive definite.}, we may argue as in the proof
of the area theorem~\cite{HE} to show that $\theta =0$.
It then also follows that $ \widehat{\sigma}_{ab}=0$, and
\bena
 \pounds_n \phi^i = 0 \,,  \quad
 V^I_a =0 \,, \quad \mbox{ on $H$}\,.
\label{condi:phi:V}
\eena
The relations $\theta = 0 = \widehat \sigma_{ab}$ on $H$ are equivalent to
\ben
  \pounds_n \gamma_{ab} =0 \,, \quad \mbox{on $H$} \,,
\een
which---when substituted into the Einstein equations eqs.~(\ref{nAR})
and (\ref{Tnp}) and combined with eqs.~(\ref{condi:phi:V})---give
\ben
D_a \alpha = \half \pounds_n \beta_a \,, \quad \mbox{on $H$}\,.
\een
In the non-extremal case, we may now argue as in~\cite{HIW07}
that we can always pass to a modified system of Gaussian
null coordinates with associated quantities $\tilde \alpha, \tilde \beta_a,
\tilde \gamma_{ab}, \tilde \phi^i, \tilde V^I_a, \tilde S^I$ etc. such that $\tilde \alpha$
is constant and non-zero over $H$.
In the extremal case, we can use the same arguments as in the previous
section to construct a modified system of Gaussian null coordinates such that
$\tilde \alpha = 0$ on $H$ under the assumption that the vector of angular
velocities $\underline \Omega$ verifies
the diophantine condition given in the statement of Theorem~1.
We assume from now on that our Gaussian null coordinates have been chosen
in this way in either case, and we drop the ``tilde" from the corresponding
quantities again to lighten the notation. Thus it follows that
\ben
 \pounds_n \beta_a = 0 \,, \quad \mbox{on $H$} \,.
\een
Next, from the Bianchi identities, eq.~(\ref{id:Bianchi})
[see eq.~(\ref{BI:uAB})], and condition, eq.~(\ref{condi:phi:V}),
we find that
\bena
\pounds_n U^I_{ab} = 0 \,, \quad \mbox{on $H$} \,.
\eena
Using conditions, eqs.~(\ref{condi:phi:V}), we immediately can show that
$n_a (\nabla_b \phi^i)F^{Iab} = 0$ on $H$. Then, using the results above
and the equation of motion for the gauge field, eq.~(\ref{eom:gauge}),
contracted with $g_{ab}n^b$, we obtain
\ben
 \pounds_n S^I = 0 \,,  \quad \mbox{on $H$} \,.
\een
At this point, we can show that
\bena
 \pounds_n \pounds_\ell \gamma_{ab} = 0 \,,  \quad \mbox{on $H$} \,.
\label{nlgamma}
\eena
Indeed, if we take a Lie derivative $\pounds_n$ of
eq.~(\ref{eom:grav}), and contract with $p^c{}_ap^d{}_b$ (see eqs.~(\ref{ABR}),
(\ref{eq:grav}), and (\ref{Tpp})), then we obtain
\ben
 \pounds_n \{\pounds_n \pounds_\ell \gamma_{ab}+ \alpha \pounds_\ell
\gamma_{ab}\} = 0 \,, \quad \mbox{on $H$} \,,
\een
as in the vacuum case. In the non-extremal case, eq.~(\ref{nlgamma})
follows from the argument below eq.~(72) of \cite{HIW07},
For the extremal case, i.e. when $\alpha=0$,
the same argument as given around eq.~(\ref{eq:nlgamma}) above applies.

\medskip
We next show that
\bena
 \pounds_n W^I_a = 0 \,, \quad \mbox{on $H$} \,.
\label{nW}
\eena
First, taking a Lie derivative $\pounds_n$ of the gauge field equation,
eq.~(\ref{eom:gauge}) and contracting with $p^c{}_a$ (see eq.~(\ref{eom:gauge:C})),
we have
\ben
 \pounds_n \pounds_n q^{ab}W^I_b + 2\alpha \pounds_n q^{ab}W^I_b
 + \pounds_n \pounds_\ell q^{ab}V^I_b = 0 \,, \quad \mbox{on $H$}\,.
\label{eq:WWV}
\een
Second, taking a Lie derivative $\pounds_n$ of the Bianchi identities,
eq.~(\ref{id:Bianchi}) (see eq.~(\ref{BI:urA})) and using $\pounds_n S^I=0$,
we have
\ben
 \pounds_n \pounds_\ell V^I_a
 - \pounds_n \pounds_n W^I_a = 0 \,, \quad \mbox{on $H$}\,.
\een
Substituting this into the above equation, eq.~(\ref{eq:WWV}), we find
\ben
  \pounds_n \{ \pounds_n W^I_a + \alpha W^I_a \}= 0 \,, \quad \mbox{on $H$}\,.
\een
Then eq.~(\ref{nW}) follows by the same type of argument as below
eq.~(72) of \cite{HIW07} for the non-extremal case $\alpha=\kappa \neq 0$, and the type of
argument as below eq.~(\ref{eq:nlgamma}) for the extremal case
$\alpha = \kappa =0$.

\medskip
Thus, we have shown that all eqs.~(\ref{cid}), (\ref{cid:s}), and
(\ref{cid4}), for $m=0$, and the first of eq.~(\ref{cid}) $m=1$ are
satisfied on $H$. The remaining equations for all other values of $m$ are
verified by the same type of inductive argument as in \cite{HIW07} for
the non-extremal case, and as in the previous section for the extremal case.

\medskip
In summary, we have verified that Theorems~1 and~2 continue to hold
in the presence of matter fields described by the above action~\eqref{action}.
In the non-extremal case, the diophantine condition stated in Theorem~1 is
not required.

\section{Discussion}\label{sect:discussion}
In this paper, we have considered degenerate (extremal) stationary black hole spacetimes with
Killing field $t^a$. We showed that, if
the vacuum Einstein equations hold and the spacetime is asymptotically flat, then
there exists a Killing field $K^a$ that is tangent and normal to the horizon generators, i.e. the black
hole horizon is a Killing horizon. We also proved that if $t^a$ is not everywhere
tangent to the null generators (so that $K^a \neq t^a$), then there exist $N$ additional
rotational Killing fields, where $N$ is at least one. Our proof relied on two technical assumptions
about the nature of the black hole: First we assumed that the spacetime metric is real analytic.
Secondly, we had to assume that the corresponding angular velocities $\underline \Omega
= (\Omega_1, \dots, \Omega_N)$ satisfy the ``diophantine condition"~\eqref{dioph}. This condition is
automatically satisfied when $N=1$, in which case the spacetime isometries generated by the timelike
Killing field map the horizon generators to themselves after the
period $T=2\pi/ \Omega_1$. However, when $N>1$---which can happen
only in $n>4$ spacetime dimensions---the
diophantine condition is non-trivial. In this sense, our
theorem is weaker than that obtained in our previous paper~\cite{HIW07} for the non-degenerate case, because no
assumption of that type had to be made there. We also considered a class of theories containing
scalar and abelian gauge fields and derived similar results in this context.

\medskip
Let us make a few elementary remarks concerning the diophantine condition~\eqref{dioph}.
First, it is well-known that this condition holds for all $\bomega = (\Omega_1, \dots, \Omega_N) \in \mr^N$
except for a set of Lebesgue measure zero\footnote{
Since we know that the orbits $t \to t\bomega \, {\rm mod} \, {\mathbb Z}^N$
are dense on ${\mathbb T}^N$, it follows that the entries of $\bomega$ are
linearly independent over $\mathbb Z$. By the Schmidt-subspace theorem~\cite{schmidt},
if there is an $i$ such that the ratios $\Omega_j/\Omega_i$ are
algebraic numbers for $j=1, \dots, N$, then $\bomega$ satisfies the diophantine
condition. Of course, the set of $\bomega$ for which this condition is
satisfied is much larger---it has full measure.}.
This follows immediately from the fact that the set
where the condition~\eqref{dioph} does {\em not} hold for any $q$ is contained in the intersection
$\cap_{q} \Lambda_q$ of the sets
\ben
\Lambda_q = \{ \bomega = (\Omega_1, \dots, \Omega_N) \in \mr^N \mid |\bomega \cdot \n| \le |\bomega | \cdot |\n|^{-q}
\quad \text{for some $\n$ with $|\n|>1$} \} \, .
\een
This intersection has Lebesgue measure zero,
\ben
{\rm measure} \left( \bigcap_{q=1}^\infty \Lambda_q \right) = 0 \, .
\een
For completeness, let us briefly show this: If $B_r$ denotes the ball of radius $r$ in $\mr^N$, then we have
\bena
{\rm measure}(\Lambda_q \cap B_r)  &\le& \sum_{|\n|>1} {\rm const.} \, r^N |\n|^{-q-1} \non \\
&\le&  {\rm const.} \, r^N
\, \int_{|\x|>1} |\x|^{-q-1} \, \dd^N \x \le \frac{{\rm const.} \, r^N}{q+1-N} \, .
\eena
Since this goes to zero for $q \to \infty$, the claim follows immediately.
Thus, it would seem that the assumptions of our theorem are satisfied in
almost the entire space of possible parameters $\underline \Omega$.
Unfortunately, our analysis gives no indication exactly what the true
parameter space of possible values of $\underline \Omega$ for $n$-dimensional
stationary black holes really is.
For example, it could still happen that
this space is very sparsely populated for certain types of black holes,
i.e., it is theoretically possible that extremal black holes could only
exist for $\underline \Omega$ in a set of measure zero.
In that case, the statement that the assumptions are satisfied for almost all
$\underline \Omega \in \mr^N$ would have little value. Let us look at the
example of a $5$-dimensional black ring. It admits
$N=2$ rotational Killing fields and there is a regular limit in which
the horizon becomes degenerate. In this limit, the
angular velocities $\underline \Omega=(\Omega_1, \Omega_2)$
are non-vanishing and satisfy $\theta = \Omega_1/\Omega_2 = \pm 1$
for the first branch of solutions, or
\bena
\theta = (1+x)/2\sqrt{x}\,, \quad 0 \le x < \infty
\eena
for the second branch (see, e.g., \cite{ElvangRodriguez08,ER08}).
Thus, for the first branch, $\theta$ is in particular rational,
and the orbits of the projection of $t^a$ onto the space of
null-generators of the horizon consequently always close.
For the second branch, $\theta$ varies continuously and may
thus be irrational. The vector $\underline \Omega$ satisfies the
diophantine condition for almost all extremal black hole solutions,
but there is a measure zero set where it does not, corresponding
to certain transcendental values of $\theta$. However, even in those
exceptional cases the black hole continues to have $N=2$ rotational Killing
fields and is a Killing horizon. This suggests that our theorem might
be true even dropping the diophantine condition.

\medskip
Secondly, as we have seen, the diophantine condition is needed
in lemma~\ref{lemma:1} to control the sizes of denominators of
the form $|\bomega \cdot \n|$ when $\n$ becomes large. It appears that this
condition cannot easily be lifted for generic analytic functions $J(x)$
in this lemma. Indeed, let us suppose that we have, say $N=2$,
and $\theta = \Omega_1/\Omega_2$ is given by the series
\ben
\theta = \sum_{i=0}^\infty \frac{1}{a_i}
\een
where $a_i$ is defined recursively by $a_0=1$ and $a_{i+1} = 2^{K a_i}$, with $K \in \mathbb N$.
This series is converging rapidly to a transcendental\footnote{By eq.~\eqref{rational}, $\theta$
cannot be rational. If $x$ were not transcendental,
then by a classic theorem of Liouville, we would have $|\theta-p/q|>{\rm const.} \, |q|^{-d}$, where
$d$ is the degree of the algebraic number $\theta$. This condition is not satisfied due to
eq.~\eqref{transcend}.}
number $1<\theta<2$, as $a_{i+1} - a_i \ge 1$ and $a_{i+1}/a_i = 2^{K(a_i-a_{i-1})} \le 2^{-K}$.
If $p_k/q_k$ denotes the $k$-th partial sum,
\ben\label{rational}
\frac{p_k}{q_k} = \frac{1 + \sum_{i=1}^k \frac{a_k}{a_{k-1}} \cdots \frac{a_i}{a_{i-1}} }{a_k} \, ,
\een
then $h.c.f.(p_k, q_k) = 1$. Furthermore, we have
\ben\label{transcend}
|\theta-p_k/q_k| = \frac{1}{a_{k+1}}\left( 1+  \sum_{i=k+1}^\infty \frac{a_{i-1}}{a_{i}} \cdots \frac{a_{k+1}}{a_{k+2}}
\right) \le
\frac{\sum_{i \ge 0} 2^{Ki}}{a_{k+1}} \le \frac{2}{a_{k+1}} =2^{-Kq_k+1} \, .
\een
This implies that an exponential type sum of the form considered in lemma~1
\ben
\sum_{p,q \in \mz} \frac{\e^{-c|p|-c|q|}}{|p/q-\theta|}
\een
cannot converge for sufficiently large $K$, as there are always terms of size at least
\ben
\frac{\e^{-c|p_k|-c|q_k|}}{|p_k/q_k-\theta|} \ge \frac{1}{2} a_{k+1} \e^{-c|a_k|} \ge 2^{a_k(K-c\log_2 \e)-1}
\ge 2^{k(K-c\log_2 \e)-1} \to \infty \quad (k \to \infty)
\een
in this sum. In the proof of lemma~1, $\e^{-c|p|-c|q|}$ bounds the Fourier coefficients~\eqref{fouriert} of $J(x)$.
If it is only known that $J(x)$ analytic, then no better bound can be derived, and the solution
to the equation $\pounds_s h(x) = J(x)$ consequently cannot be obtained
by the method of the lemma. However, in our case $J(x) = 1 - \e^{-F(x)}$, where
$F$ in turn satisfies $\pounds_s F = \alpha$. It might be possible that
further constraints can be derived on the Fourier coefficients of $J(x)$ from
such a relation combined with Einstein's equations. But we have not been able to find such relations.

\medskip
Let us finally make a remark about the assumption of analyticity.
It is known that the Einstein-Maxwell system admits extremal
multi-black hole solutions which have non-smooth---hence
non-analytic---horizons \cite{GHT95,Welch95,CR07}.
Therefore when including Maxwell fields,
the analyticity assumption---which is one of the key assumptions
in our proof---is not entirely plausible. As shown
in \cite{HIW07,FRW99,Racz00}, the analyticity assumption
can be partially removed
%\footnote{%%%
%See for an attempt to fully remove the analyticity assumption
%in the entire spacetime region in $4$-dimensions \cite{IK07}.
%}%%%
when the event horizon is non-degenerate. In that case,
the horizon can be shown to be isometric to a portion of a
bifurcate null hypersurface \cite{RW92,RW96}, and one can use
the characteristic initial value formulation for Einstein's
equations~\cite{Friedrich91,MzH90,Rendall90} on the bifurcate
null hypersurface in order to extend $K^a$ defined on the horizon
to the black hole interior region.
This is, however, not the case for degenerate horizons, since
on such a horizon with $\kappa=0$, the completeness of
the Killing parameter of $K^a$ on the horizon implies that
the horizon generator is affinely complete and hence
there is no bifurcate surface. Thus, the key issue
when generalizing our results to the Einstein-Maxwell
system is whether the diophantine condition holds, and
whether the solutions is analytic, including a neighborhood
of the horizon.

\medskip
An interesting generalization of this work would be to consider vacuum
spacetimes which are not asymptotically flat in the standard sense
(with asymptotic infinity of topology $S^{n-2}$), but instead for example asymptotically Kaluza-Klein,
with asymptotic infinity of the form $S^2 \times Y$, with $Y$ a compact manifold of dimension $n-4$.
In the non-degenerate case, there would be no change in our analysis of the local horizon geometry, and we could construct
a vector field $K^a$ in a neighborhood of the horizon $H$ satisfying~\eqref{civ1}, i.e., the Killing
equation $\pounds_K g_{ab}=0$ holds on $H$ to all orders in a Taylor expansion off of $H$. The same would also
apply in the degenerate case if we assume a diophantine condition~\eqref{dioph}. However, in both cases
it might no longer be possible to construct the desired $K^a$ {\em globally} by analytic continuation:
The point is that we are only guaranteed to get a single-valued extension if the exterior part of the spacetime
is simply connected~\cite{Nomizu60}. Now, the topological censorship theorem~\cite{Galloway01,Galloway99} guarantees that
\ben
\pi_1(M_{\rm exterior}) \cong \pi_1 (S^2 \times Y)
\een
but unlike in the case of an asymptotically flat spacetime with infinity $S^{n-2}$,
the fundamental group $\pi_1(S^2 \times Y)$ no longer need to vanish. Nevertheless, if
$\pi_1(Y) = 0$, then it does, and we presumably again get results analogous to Theorems~1 and~2.

\medskip
\noindent
\paragraph{\bf Acknowledgements:} We would like to thank P. Chrusciel,
H. Reall, and Bob Wald for valuable discussions and comments. SH
would like to thank M. Huxley for discussions on the diophantine
condition. AI wishes to thank School of Mathematics, Cardiff University
for its hospitality during the time some of this research was carried out
and also wishes to thank the Perimeter Institute for Theoretical Physics
for its hospitality during the time other parts of this research were
carried out. This research was supported in part by Perimeter Institute
for Theoretical Physics.

\appendix %----------------

\section{Ricci tensor in Gaussian null coordinates}\label{sect:A}

In this Appendix, we provide expressions for the Ricci tensor in a
Gaussian null coordinate system. As derived in section~\ref{sect:2},
in Gaussian null coordinates, the metric takes the form
\ben
g_{ab} = 2 \left(
                 \nabla_{(a} r - r \alpha \nabla_{(a}u - r \beta_{(a}
           \right) \nabla_{b)}u
       + \gamma_{ab} \,,
\een
where the tensor fields $\beta_a$ and
$\gamma_{ab}$ are orthogonal to $n^a$ and $\ell^a$.
The horizon, $H$, corresponds to the surface $r=0$. We note
that ${\gamma^a}_b$ is the orthogonal
projector onto the subspace of the tangent
space orthogonal to $n^a$ and $\ell^a$, and that when $r \beta_a \neq 0$,
it differs from the orthogonal
projector, ${q^a}_b$, onto the surfaces $\Sigma(u,r)$.
It is worth noting that in terms of the Gaussian null coordinate components
of $\gamma_{ab}$, we have
$q^{ab} =
(\gamma^{-1})^{AB} (\partial /\partial x^A)^a (\partial/\partial x^B)^b$.
It also is convenient
to introduce the non-orthogonal projector ${p^a}_b$, uniquely
defined by the conditions that ${p^a}_b n^b = {p^a}_b \ell^b = 0$ and that
${p^a}_b$ be the identity map on vectors that are tangent to $\Sigma(u,r)$.
The relationship between ${p^a}_b$ and ${\gamma^a}_b$ is given by
\ben
{p^a}_b = -r \ell^a \beta_b + {\gamma^a}_b \,.
\een
In terms of Gaussian null coordinates, we have
$p^a{}_b = (\partial/\partial x^A)^a (\dd x^A)_b$, from which it is easily
seen that
$\pounds_n p^a{}_b = 0 = \pounds_\ell p^a{}_b$. It also is easily seen that
$q^{ac} \gamma_{cb} = p^a{}_b$ and that $p^a{}_b q^b{}_c = q^a{}_c$.

\medskip
We define the derivative operator $D_c$ acting on a tensor
field $T^{a_1 \dots a_r}{}_{b_1 \dots b_s}$ by the following prescription.
First, we project the indices of the tensor field by $q^a{}_b$, then we
apply the covariant derivative $\nabla_c$, and we then again project
all indices using $q^a{}_b$. For tensor fields intrinsic to $\Sigma$, this
corresponds to the derivative operator associated with the metric $q_{ab}$.
We denote the Riemann and Ricci tensors associated with $q_{ab}$ as
${\mathcal R}_{abc}{}^d$ and ${\mathcal R}_{ab}$.

\medskip
The Ricci tensor of $g_{ab}$ can then be written in the following form:
\bena
\label{nnR}
 n^an^bR_{ab} &=&
 -\half q^{ab} \pounds_n\pounds_n \gamma_{ab}
 +\quater q^{ca}q^{db}(\pounds_n \gamma_{ab})\pounds_n \gamma_{cd}
 +\frac{1}{2} \alpha \: q^{ab}\pounds_n \gamma_{ab}
\non \\
 &+&
 \frac{r}{2} \cdot
 \Bigg[\;
       4 \alpha \pounds_\ell \pounds_\ell \alpha
       + 8 \alpha \pounds_\ell\alpha
       + (\pounds_\ell\alpha) q^{ab}\pounds_n\gamma_{ab}
\non \\
 && \qquad \,
      + q^{ab}\pounds_\ell\gamma_{ab} \cdot
                \Big\{
                      -\pounds_n \alpha
                      -r q^{cd}\beta_c\pounds_n\beta_d
\non \\
 && \qquad \qquad \qquad \qquad \,\,\:
                      +(rq^{cd}\beta_c\beta_d+2\alpha)\pounds_\ell(r\alpha)
                      + rq^{cd}\beta_c D_d \alpha
                \Big\}
\non \\
 && \qquad \,
     + 2q^{ab}{D}_a
          \left\{
             \beta_b\pounds_\ell(r\alpha) + {D}_b\alpha-\pounds_n\beta_b
          \right\}
\non \\
 && \qquad \,
     +q^{bc}\pounds_\ell(r\beta_c)\cdot
          \Big\{
                 (rq^{ef}\beta_e\beta_f+2\alpha) \pounds_\ell(r\beta_b)
\non \\
 && \qquad \qquad \qquad \qquad \quad
                - 4 D_b\alpha
                + 2 \pounds_n\beta_b + 4rq^{ae}\beta_e {D}_{[a}\beta_{b]}
          \Big\}
\non \\ && \qquad \,
     + 2(\pounds_\ell\alpha)\pounds_\ell(r^2q^{ab}\beta_a\beta_b)
     + 4rq^{ab}\beta_a\beta_b \pounds_\ell\alpha
     + 2 rq^{ab}\beta_a\beta_b \pounds_\ell \pounds_\ell \alpha
\non \\ && \qquad \,
     +2 q^{ab}\beta_a\pounds_\ell(r\beta_b)\cdot
      \left\{
             2 \pounds_\ell(r\alpha)
            - \half r q^{cd}\beta_c\pounds_\ell(r\beta_d)
      \right\}
\non \\
 && \qquad \,
     + 2r^{-1}\pounds_\ell \left\{
                                 r^2q^{ab}\beta_a(D_b\alpha-\pounds_n\beta_b)
                           \right\}
     + 2r^{-1}\alpha \pounds_\ell(r^2 q^{ab}\beta_a\beta_b)
 \Bigg] \,,
\eena

\bena
\label{nlR}
 n^a\ell^bR_{ab}
     &=& -2 \pounds_\ell\alpha
         + \quater q^{ca}q^{db}(\pounds_n\gamma_{cd})\pounds_\ell\gamma_{ab}
         -\half q^{ab} \pounds_\ell \pounds_n \gamma_{ab}
         -\frac{1}{2} \alpha\: q^{ab} \pounds_\ell\gamma_{ab}
         -\half q^{ab}\beta_a\beta_b
\non \\
 &+&
   \frac{r}{2}\cdot
    \Bigg[
          - 2\pounds_\ell \pounds_\ell\alpha
          -\half q^{ab}\pounds_\ell\gamma_{ab} \cdot
           \left\{
                  2\pounds_\ell\alpha + q^{cd}\beta_c\pounds_\ell(r\beta_d)
           \right\}
\non \\
 && \qquad \, \,
          - q^{ab}\beta_a\pounds_\ell\beta_b
          - \pounds_\ell\{ q^{ab}\beta_a\pounds_\ell(r\beta_b)\}
          - q^{ab}D_a(\pounds_\ell\beta_b)
    \Bigg] \,,
\eena

\bena
\label{nAR}
 n^bp^c{}_aR_{bc} &=&
  - p^b{}_aD_b \alpha
  + \half\pounds_n\beta_a
  + \quater \beta_a q^{bc} \pounds_n \gamma_{bc}
  - p^d{}_{[a}p^e{}_{b]}D_d(q^{bc}\pounds_n \gamma_{ce})
\non \\
&+&
   \frac{r}{2}\cdot
   \Bigg[\;
         \half (q^{bc}\pounds_n\gamma_{bc})\pounds_\ell \beta_a
         + \pounds_n\pounds_\ell\beta_a
         + 2\alpha \pounds_\ell\beta_a
\non \\
 && \qquad
         + \pounds_\ell(r\beta_a) \cdot
                       \left\{
                              r^{-1}\pounds_\ell(r^2q^{bc}\beta_b\beta_c)
                              + 2 \pounds_\ell\alpha
                       \right\}
\non \\
 && \qquad
         - 2p^b{}_aD_b(\pounds_\ell\alpha)
         + \pounds_\ell(q^{bc}\beta_b\pounds_n\gamma_{ca})
         - 2r^{-1}\pounds_\ell \left(
                                     r^2 q^{cd}\beta_cp^b{}_aD_{[b}\beta_{d]}
                               \right)
\non \\
 && \qquad
         - \half q^{bc}\pounds_\ell\gamma_{bc}\cdot
           \Big\{
                  - (rq^{ef}\beta_e\beta_f+2\alpha)\pounds_\ell(r\beta_a)
\non \\
&& \qquad \qquad \qquad \qquad \qquad
                  + 2 p^d{}_aD_d\alpha
                  - q^{bc}\beta_b\pounds_n \gamma_{ca}
                  + 2r q^{ef}\beta_ep^d{}_aD_{[d}\beta_{f]}
           \Big\}
\non \\
 && \qquad
         - 2 \pounds_\ell(\alpha \beta_a)
         - 2 r(\pounds_\ell \alpha) \pounds_\ell \beta_a
         + p^d{}_aD_b\left\{q^{bc}\beta_c\pounds_\ell(r\beta_d) \right\}
\non \\
 && \qquad
         - 2 p^b{}_a q^{cd}D_dD_{[b}\beta_{c]}
         - q^{bc} (\pounds_\ell\beta_b) \pounds_n \gamma_{ca}
\non \\
 && \qquad
         - q^{bc}\pounds_\ell(r\beta_b)\cdot
           \Big\{
                  (rq^{ef}\beta_e\beta_f+2\alpha) \pounds_\ell\gamma_{ca}
                  + p^d{}_a D_c \beta_d
\non \\
&& \qquad \qquad \qquad \qquad \qquad
                  + \beta_c\pounds_\ell(r\beta_a)
                  - rq^{ef}\beta_c\beta_f\pounds_\ell\gamma_{ea}
           \Big\}
\non \\
 && \qquad
          + q^{bc}(\pounds_\ell\gamma_{ca})\cdot
             \left\{
                    2 \beta_b\pounds_\ell(r\alpha)
                    + 2 D_b \alpha
                    - \pounds_n \beta_b
                    + 2r q^{de}\beta_eD_{[b}\beta_{d]}
             \right\}
   \Bigg]\,,
\eena

\bena
\label{llR}
 \ell^a \ell^b R_{ab}
        &=& - \half q^{ab} \pounds_\ell \pounds_\ell \gamma_{ab}
            + \quater q^{ca}q^{db}
              (\pounds_\ell\gamma_{ab})\pounds_\ell\gamma_{cd} \,,
\eena

\bena
\label{lAR}
 \ell^bp^c{}_a R_{bc}
      &=&
         - \quater \beta_a q^{bc}\pounds_\ell \gamma_{bc}
         - \pounds_\ell \beta_a
         + \half q^{bc}\beta_c\pounds_\ell \gamma_{ab}
         - p^d{}_{[a} p^e{}_{b]}D_d
           \left(q^{bc}\pounds_\ell \gamma_{ce} \right)
\non \\
 & + &  \frac{r}{2}\cdot
        \Bigg[\;
                - \pounds_\ell \pounds_\ell \beta_a
               + \pounds_\ell
                \left(q^{bc}\beta_c\pounds_\ell\gamma_{ab} \right)
\non \\
&& \qquad \qquad
                + \half (q^{cd}\pounds_\ell \gamma_{cd})
                    \left(
                          - \pounds_\ell\beta_a
                          + q^{be}\beta_e\pounds_\ell \gamma_{ab}
                    \right)
        \Bigg] \,,
\eena

\bena
\label{ABR}
p^c{}_ap^d{}_b R_{cd}
        &=&
         - \pounds_\ell\pounds_n \gamma_{ab}
         - \alpha \pounds_\ell \gamma_{ab}
         + p^c{}_ap^d{}_b {\cal R}_{cd}
         - p^c{}_{(a} p^d{}_{b)} D_c\beta_d
         - \half \beta_a\beta_b
\non \\
        &+&
           q^{cd} \left(\pounds_\ell\gamma_{d(a}\right)\pounds_n \gamma_{b)c}
         - \quater
           \left\{
                   (q^{cd}\pounds_n \gamma_{cd})\pounds_\ell \gamma_{ab}
                 + (q^{cd}\pounds_\ell \gamma_{cd})\pounds_n \gamma_{ab}
           \right\}
\non \\
 &+& \frac{r}{2} \cdot
     \Bigg[
           - 2\alpha \pounds_\ell \pounds_\ell \gamma_{ab}
           - p^e{}_ap^f{}_bD_c(q^{cd}\beta_d\pounds_\ell \gamma_{ef})
\non \\
  && \qquad \,
           - \half (q^{cd}\pounds_\ell\gamma_{cd})
             \left\{
                     (rq^{ef}\beta_e\beta_f+2\alpha)\pounds_\ell\gamma_{ab}
                   + 2 p^e{}_{(a} p^f{}_{b)} D_e\beta_f
             \right\}
\non \\
  && \qquad \,
           - 2(\pounds_\ell \alpha) \pounds_\ell \gamma_{ab}
           - r^{-1}\{\pounds_\ell(r^2q^{ef}\beta_e\beta_f)\}
                   \pounds_\ell \gamma_{ab}
\non \\
  && \qquad \,
           - rq^{ef}\beta_e\beta_f \pounds_\ell \pounds_\ell \gamma_{ab}
           - 2\pounds_\ell \{ p^c{}_{(a} p^d{}_{b)} D_c\beta_d \}
\non \\
  && \qquad \,
           - 2\beta_{(a}\pounds_\ell\beta_{b)}
           - r(\pounds_\ell\beta_a) \pounds_\ell\beta_b
           - rq^{ce}q^{df}\beta_c\beta_d
             (\pounds_\ell \gamma_{ae})\pounds_\ell\gamma_{bf}
\non \\
  && \qquad \,
           + 2q^{cd}\beta_d
             \left\{\pounds_\ell(r\beta_{(a}) \right\}\pounds_\ell\gamma_{b)c}
           + 2p^e{}_{(a}p^f{}_{b)}q^{cd}\left(D_d \beta_e \right)
             \pounds_\ell\gamma_{fc}
\non \\
  && \qquad \,
           + q^{cd} (r q^{ef}\beta_e\beta_f+2\alpha)
                      (\pounds_\ell\gamma_{ca})\pounds_\ell\gamma_{db} \,
     \Bigg] \,.
\eena

\section{Gravity coupled to matter fields} \label{sect:B}
We start from the action
\bena
 S = \int \dd^n x \sqrt{- g}
\Big(
  R -\half f_{ij}(\phi) g^{ab} \nabla_{a} \phi^i \nabla_b \phi^j
    - U(\phi)
    - \frac{1}{4} h_{IJ}(\phi)
      g^{ac} g^{bd} F^I_{ab} F^J_{cd}
\Big)
%+ S_{\rm CS}
\,,
\eena
where $F^I_{ab} = \nabla_a A^I_b - \nabla_b A^J_a $,
and where $f_{ij}(\phi)$ and $h_{IJ}(\phi)$ are positive definite metrics
on the spaces of scalar fields, $\phi^i$, and gauge fields, $A^I_a$,
respectively. Variation of $S$ gives
\bena
% \bullet
&& R_{ab}
  = T_{ab}-\frac{T}{n-2}g_{ab} \,,
% \non
% \\
% && \quad
% = f_{ij} (\phi) \nabla_a \phi^i \nabla_b \phi^j
%   + h_{IJ}(\phi) g^{cd} F^I_{ac}F^J_{bd}
%   + \frac{2}{n-2}g_{ab}
%     \left[
%        U(\phi) - \frac{1}{4}h_{IJ}(\phi)F^I_{cd}F^{J cd}
%     \right] \,,
\label{eq:grav}
\\
&&
\non \\
% \bullet
&&
  \nabla_a (f_{ij} (\phi) \nabla^a \phi^j)
 - \half f_{jk |i} \nabla^a \phi^j \nabla_a \phi^k
 - U_{| i}
 - \frac{1}{4} h_{IJ|i}F^I_{ab}F^{J ab} = 0 \,,
\label{eq:scalar}
\\
&&
\non
\\
% \bullet
&&
 \nabla_c\left( h_{IJ}(\phi) F^{J ca} \right) =0 \,,
\label{eq:maxwell}
\eena
where the stress-energy tensor, $T_{ab}$, is given by
\bena
 T_{ab} &=& f_{ij}(\phi)\nabla_a\phi^i \nabla_b \phi^j
            - \half g_{ab}
              \Big\{
                   f_{ij}(\phi)\nabla^c \phi^i \nabla_c \phi^j
                   + 2U(\phi)
              \Big\}
\non \\
 && \qquad \quad
       + h_{IJ}(\phi)
              \Big\{
                    g^{cd} F^I_{ac} F^J_{bd}
                   - \frac{1}{4} g_{ab}
                     F^I_{cd} F^{J cd}
              \Big\} \,,
\label{def:Tab}
\eena
and $T=T^c{}_c$, and where here and in the following
the vertical bar denotes the derivative with respect to a scalar field,
$\phi^i$, e.g., $f_{jk |i}= \partial f_{jk}(\phi)/ \partial \phi^i$.
% and
% \bena
% && T_{ab} - \frac{T}{n-2}g_{ab}
%  = f_{ij}(\phi)\nabla_a \phi^i \nabla_b \phi^j
%    + h_{IJ}(\phi)g^{cd} F^I_{a c} F^J_{bd}
% \non \\
% && \qquad \qquad \qquad \qquad
%    + \frac{2}{D-2}g_{ab}
%               \Big\{
%                     U(\phi)
%                    - \frac{1}{4}h_{IJ}(\phi)
%                      F^I_{cd} F^{J cd}
%               \Big\} \,,
% \eena
% where
% \bena
% \frac{1}{4}h_{IJ}(\phi) F^I_{ab} F^{J ab}
% &=&
%   h_{IJ}(\phi) \cdot
%  \Bigg\{
%        - \half S^IS^J
%        + \gamma^{AB}V_A^I W_B^J
%        + \frac{1}{4}\gamma^{AB}\gamma^{CD} U_{CA}^I U_{DB}^J
%  \Bigg\}
% \non \\
% &&
%  + h_{IJ}(\phi) \cdot r \cdot
%  \Bigg[
%        \beta^A W^I_A S^J
%        - 2 \beta^C \gamma^{AB}  W_A^I U_{BC}^J
%        + \alpha \gamma^{AB} W_{A}^I W_{B}^J
%  \Bigg]
% \non \\
% &&
%  + \half h_{IJ}(\phi) \cdot {r^2} \cdot
% \Bigg[
%       \left( \beta^2 \gamma^{AB} - \beta^A \beta^B \right) W_A^I W_B^J
% \Bigg] \,.
% \eena
%
% Let us denote
% \ben
% {\cal T}_{ab} = T_{ab}- \frac{T^c{}_c}{n-2}g_{ab} \,.
% \een
In terms of the tensor fields, $S^I,\:V^I_a,\:W^I_a\:U^I_{ab}$,
and the metric variables $\alpha, \beta_a, \gamma_{ab}$ introduced in the context of Gaussian null
coordinates, the right-hand side of eq.~(\ref{eq:grav}) can be decomposed as follows:
\bena
\label{Tnn}
% n^a n^b {\cal T}_{ab}
 n^an^b\left( T_{ab} - \frac{T}{n-2}g_{ab} \right)
 &=& f_{ij}(\phi) (\pounds_n \phi^i) \pounds_n\phi^j
\non \\
 &&
  + h_{IJ}(\phi)\cdot
    \Bigg\{
           q^{ab}V^I_a V^J_b
         + 2 {r}\cdot
           \Big(
                \alpha S^IS^J + \beta^a V^I_a S^J
           \Big)
         + {r^2} \cdot \beta^b\beta_b S^IS^J
    \Bigg\}
 \non \\
 &&
  - r \cdot \frac{4 \alpha}{n-2} \cdot {\cal T} \,,
\\
\label{Tnl}
 n^a \ell^b \left( T_{ab} - \frac{T}{n-2}g_{ab} \right)
 &=& f_{ij}(\phi) (\pounds_n \phi^i ) \pounds_\ell \phi^j
\non \\
 &&
  + h_{IJ}(\phi)\cdot
    \Bigg\{
           q^{ab} V^I_aW^J_b
         + {r} \cdot \beta^a W^I_a S^J
    \Bigg\}
% \non \\
%  &&
  + \frac{2}{n-2} \cdot {\cal T} \,,
\\
\label{Tnp}
 n^b p^c{}_a \left( T_{bc} - \frac{T}{n-2}g_{bc} \right)
 &=& f_{ij}(\phi) (\pounds_n \phi^i) p^b{}_a {D}_b \phi^j
\non \\
 &&
  + h_{IJ}(\phi)\cdot
    \Bigg\{
           - S^I V^J_a
           + q^{bc}V^I_bU^J_{ac}
\non \\
&& \qquad \qquad \quad
           + {r}\cdot
           \Big[
                - 2\alpha S^I W^J_a + \beta^b U^I_{ab}S^J
                - \beta^b V^I_b W^J _a
           \Big]
\non \\
&& \qquad \qquad \quad
           - {r^2} \cdot \beta^c\beta_c S^IW^J_a \:
    \Bigg\}
% \non \\
%  &&
    - r\cdot \frac{2 \beta_a}{n-2} \cdot {\cal T} \,,
\\
\label{Tll}
 \ell^a \ell^b \left( T_{ab} - \frac{T}{n-2}g_{ab} \right)
 &=& f_{ij}(\phi) ( \pounds_\ell \phi^i) \pounds_\ell \phi^j
  + h_{IJ}(\phi) \cdot q^{ab} W^I_a W^J_b  \,,
\\
\label{Tlp}
 \ell^b p^c{}_a \left( T_{bc} - \frac{T}{n-2}g_{bc} \right)
 &=& f_{ij}(\phi) p^c{}_a (\pounds_\ell \phi^i) {D}_c \phi^j
\non \\
 && + h_{IJ}(\phi)\cdot
        \Bigg\{
               S^I W^J_a
               + q^{bc}W_b^IU_{ac}^J
               - {r} \cdot \beta^b W_b^I W_a^J
        \Bigg\}
\\
\label{Tpp}
 p^c{}_a p^d{}_b \left( T_{cd} - \frac{T}{n-2}g_{cd} \right)
 &=& f_{ij}(\phi) p^c{}_a p^d{}_b ({D}_c \phi^i) {D}_d \phi^j
\non \\
 &&
    + g_{IJ}(\phi)\cdot
        \Bigg\{ \:
                2 V^I_{(a} W^J_{b)}
               + q^{cd}U^I_{ca} U^J_{db}
\non \\
&& \qquad \qquad \quad
         + {r}\cdot
           \Big[\:
                2 \alpha W^I_aW^J_b
                - \beta^c (W^I_a U^J_{bc} +  W^I_b U^J_{ac} ) \:
           \Big]
\non \\
&& \qquad \qquad \quad
         + {r^2} \cdot \beta^c\beta_c W^I_a W^J_b \:
        \Bigg\}
% \non \\
% &&
 + \frac{2}{n-2} \cdot \gamma_{ab} \cdot {\cal T} \,,
\eena
where
\bena
% \Big[ {} \qquad {} \Big]
%  &=& U(\phi) - \frac{1}{4}h_{IJ}(\phi) F^I_{\alpha \beta} F^{J \alpha \beta}
% \non \\
{\cal T}
 &=&
 U(\phi)
 - h_{IJ} (\phi) \cdot
 \Bigg(
        - \half S^IS^J
        + q^{ab}V^I_a W^J_b
        + \frac{1}{4} q^{ab} q^{cd} U_{ca}^I U_{db}^J
 \: \Bigg)
\non \\
&&
 - r \cdot h_{IJ}(\phi) \cdot
 \Bigg(\:
        \beta^a W^I_a S^J
        - 2 \beta^c q^{ab} W^I_a U^J_{bc}
        + \alpha q^{ab} W^I_{a} W^J_{b}
 \:\Bigg)
\non \\
&&
 - \frac{r^2}{2} \cdot h_{IJ}(\phi) \cdot
 \Bigg\{\:
       \left( \beta^c\beta_c q^{ab} - \beta^a \beta^b \right) W_a^I W_b^J
 \:\Bigg\} \,.
\eena
The tensors $q^{ab}$ and $p^a{}_b$ are defined in Appendix~\ref{sect:A}.

\medskip
% \subsection{Equation for scalar fields}
Similarly, the equation for scalar fields, eq.~(\ref{eq:scalar}), are
explicitly written as
\bena
0 &=& U_{|i}
\non \\
&&
   + h_{IJ| i}
     \Bigg\{
           - \half S^IS^J + q^{ab}V_a^IW_b^J
           + \frac{1}{4} q^{ab} q^{cd}U_{ca}U_{db}
\non \\
&& \qquad \qquad \qquad
        + r\cdot
           \left(
                 \beta^a W_a^IS^J - \beta^c q^{ab}W_a^IU_{bc}^J
               + \alpha q^{ab} W_a^IW_b^J
           \right)
\non \\
&& \qquad \qquad \qquad
        + \frac{r^2}{2}\cdot
           \left[
              \left(\beta^c\beta_c q^{ab}- \beta^a\beta^b \right) W_a^IW_b^J
           \right] \:
     \Bigg\}
\non \\
&&
 - f_{ij}\cdot
   \Bigg[\: 2 \pounds_n \pounds_\ell \phi^j
     + \half (q^{ab} \pounds_n \gamma_{ab}) \pounds_\ell \phi^j
     + \half (q^{ab} \pounds_\ell \gamma_{ab}) \pounds_n \phi^j
\non \\
&& \qquad \quad \quad
       + \half (q^{ab} \pounds_\ell \gamma_{ab})
         \Big\{
              r q^{ab}\beta_a {D}_b\phi^j
              + \left(
                      2r \alpha + r^2 \beta^c\beta_c
                \right)( \pounds_\ell \phi^j)
         \Big\}
\non \\
&& \qquad \quad \quad
       + (2r\alpha + r^2 \beta^c\beta_c) (\pounds_\ell \pounds_\ell \phi^j)
       + \left\{
          2\alpha + 2r(\beta^c\beta_c + \pounds_\ell \alpha )
                  + r^2 \pounds_\ell (\beta^c\beta_c)
         \right\} ( \pounds_\ell \phi^j)
\non \\
&& \qquad \quad \quad
       + r q^{ab} \beta_a {D}_b ( \pounds_\ell \phi^j)
       + \left\{
           q^{ab} \beta_b + r \pounds_\ell (q^{ab} \beta_b)
         \right\} p^c{}_a{D}_c \phi^j
\non \\
&& \qquad \quad \quad
       + r {D}_a (q^{ab}\beta_b \pounds_\ell \phi^j)
       + q^{ab}{D}_aD_b \phi^j \:
   \Bigg]
\non \\
&&
 + \left( \half f_{jk|i} - f_{ki|j}\right) \cdot
    \Bigg[
           2 (\pounds_n \phi^j) \pounds_\ell \phi^k
       + (2r\alpha + r^2 \beta^c\beta_c)
         (\pounds_\ell \phi^j) \pounds_\ell \phi^k
\non \\
&& \qquad \qquad \qquad \qquad \qquad \qquad \quad
         + 2r q^{ab} \beta_a (\pounds_\ell \phi^j) {D}_b \phi^k
         + q^{ab} ({D}_a \phi^j) {D}_b \phi^k \:
    \Bigg] \,.
\eena

\medskip
% \subsection{Equation for gauge fields}
% \subsubsection*{Component $\nu = u$: $\ell_a \nabla_c (h_{IJ}F^{Jca}) = 0 $}
The equations of motion for the gauge fields,
$F^I_{ab}$, are given by
\bena
 0 &=& h_{IJ|i}
      \Bigg\{
              S^J ( \pounds_\ell \phi^i)
             - q^{ab}({D}_a \phi^i) W_b^J
      \Bigg\}
\non
\\
&& \qquad
    + h_{IJ}
      \Bigg\{
             - {D}_a (q^{ab} W^J_b)
             + \pounds_\ell S^J
             - \beta^a W^J_a
             + \half S^J (q^{ab} \pounds_\ell \gamma_{ab})
      \Bigg\}
\non \\
&+& {r} \cdot
  \Bigg[
       - h_{IJ|i}( \pounds_\ell \phi^i)\beta^a W^J_a
       + h_{IJ}
       \left\{
              - \pounds_\ell (\beta^a W^J_a)
              + \half \beta^c W^J_c (q^{ab} \pounds_\ell \gamma_{ab})
       \right\} \:
  \Bigg] \,,
\label{eom:gauge:u}
\eena
%
% \subsubsection*{Component $\nu = r$:
% $(\dd r)_a \nabla_c (h_{IJ} F^{Jca}) = 0$}
\bena
0 &=& h_{IJ|i}
      \Bigg\{
             - (\pounds_n \phi^i) S^J
             - q^{ab}({D}_a \phi^i) V^J_b
      \Bigg\}
\non
\\
&& \qquad \qquad
    + h_{IJ}
      \Bigg\{
             - \pounds_n S^J
             - \half S^J (q^{ab} \pounds_n \gamma_{ab})
             - {D}_a (q^{ab}V^J_b)
      \Bigg\}
\non \\
&+& r\cdot
  \Bigg[
       h_{IJ|i} (\pounds_n \phi^i)\beta^a W^J_a
     - h_{IJ|i}(D_a \phi^i)q^{ab}
       \left(
              \beta_b S^J - \beta^c U^J_{bc} + 2 \alpha W^J_b
       \right)
\non \\
&& \qquad \qquad
      + h_{IJ}
        \Big\{
             \pounds_n(\beta^a W^J_a)
            + \half (\beta^c W^J_c) (q^{ab} \pounds_n \gamma_{ab})
\non \\
&& \qquad \qquad \qquad \qquad
            -  {D}_a(q^{ab}\beta_b S^J)
            +  q^{ab}{D}_a(\beta^{c} U^J_{bc})
            - 2 {D}_a (\alpha q^{ab}W^J_b)
       \Big\} \:
  \Bigg]
\non \\
&+&
 {r^2}\cdot
 \Bigg[
       - h_{IJ|i} (D_a \phi^i) q^{ab}
           \left(
                 \beta^c\beta_c W^J_b - \beta_b \beta^c W^J_c
           \right)
\non \\
&& \qquad \qquad
       + h_{IJ} {D}_a
           \left(
                 \beta^c \beta_c q^{ab}W^J_b - q^{ab}\beta_b \beta^c W^J_c
           \right) \:
 \Bigg] \,,
\label{eom:gauge:r}
\eena
% \subsubsection*{Component $\nu = C$:
% $p^a{}_b \nabla_c (h_{IJ} F^{Jcb})=0$}
\bena
0 &=&
  h_{IJ}
  \Bigg\{
         \pounds_n (q^{ab}W^J_b)
      + \half q^{ad}W^J_d (q^{bc} \pounds_n \gamma_{bc})
      + {D}_c(q^{bc}q^{ad} U^J_{bd})
 \non \\
 && \qquad \quad
      + q^{ab}\beta_b S^J
      -  q^{ab} \beta^c U^J_{bc}
      + 2 \alpha q^{ab}W^J_b
% \non \\
% && \qquad \qquad
      + \pounds_\ell(q^{ab}V^J_b)
      + \half q^{ab}V^J_b (q^{cd} \pounds_\ell \gamma_{cd})
  \Bigg\}
\non \\
 &+&
  h_{IJ|i}
  \Bigg\{
          (\pounds_n \phi^i) q^{ab}W^J_b
        + (\pounds_\ell \phi^i) q^{ab}V^J_b
        + q^{ad}q^{bc} ({D}_b \phi^i) U^J_{cd}
  \Bigg\}
\non \\
&+& r\cdot
  \Bigg[\:
       h_{IJ}
       \Big\{
              2{D}_c(q^{b[c} q^{a]d} \beta_b W^J_d)
            + \pounds_\ell (q^{ab}\beta_b S^J)
            - \pounds_\ell (q^{ab} \beta^c U^J_{bc})
\non \\
&& \qquad \qquad \qquad \qquad
            + 2 \pounds_\ell (\alpha q^{ab} W^J_b)
            + 2 \beta^c\beta_c q^{ab}W^J_b
            - 2 \beta^a \beta^b W^J_b
     \: \Big\}
\non \\
&& \qquad \qquad
    + \Big\{\:
             \half h_{IJ} (q^{de} \pounds_\ell \gamma_{de})
           + h_{IJ|i}(\pounds_\ell \phi^i)
    \: \Big\}
              \cdot
              \left(
                    q^{ab}\beta_b S^J
                    - q^{ab}\beta^c U^J_{bc} + 2 \alpha q^{ab}W^J_b
              \right)
\non \\
&& \qquad \qquad
    + 2 h_{IJ|i} q^{c[a}q^{b]d} ({D}_b \phi^i)\beta_c W^J_d \:
  \Bigg]
\non \\
&+&
 {r^2}\cdot
 \Bigg[
       h_{IJ}
       \pounds_\ell\left(
                         \beta^c\beta_c q^{ab}W^J_b
                         - q^{ab} \beta_b\beta^cW^J_c
                   \right)
\non \\
&& \qquad \qquad
    + \Big\{
            \half h_{IJ} (q^{de} \pounds_\ell \gamma_{de})
          + h_{IJ|i}(\pounds_\ell \phi^i)
      \Big\}
              \cdot
       q^{ab}\beta^c \left(
                           \beta_c W^J_b- \beta_b W^J_c
                     \right)\:
 \Bigg] \,.
\label{eom:gauge:C}
\eena

\medskip
% \subsubsection{Bianchi equations}
The Bianchi identities $\nabla_{[a}F^I_{bc]} =0$, are written as
\bena
 \pounds_n W^I_a - \pounds_\ell V^I_a + p^c{}_aD_cS^I &=& 0 \,,
\label{BI:urA}
\\
 \pounds_n U^I_{ab} - 2p^c{}_{[a} p^d{}_{b]} D_cV^I_d &=& 0 \,,
\label{BI:uAB}
\\
 \pounds_\ell U^I_{ab} - 2p^c{}_{[a}p^d{}_{b]} D_cW^I_d &=& 0 \,,
\label{BI:rAB}
\\
 p^d{}_{[a}p^e{}_bp^f{}_{c]}D_dU^I_{ef} &=& 0 \,.
\label{BI:ABC}
\eena
% which correspond, respectively, to the components
% $(\mu,\nu,\lambda)=(u,r,A), (u,A,B), (r,A,B), (A,B,C)$.

\section{Chern-Simons theories in $n=5$} \label{sect:appe:CS}
Here we outline how the rigidity theorem can be proved
in the presence of an additional Chern-Simons term in the action.
For simplicity and concreteness, we restrict attention
to the example of minimal supergravity theory in $n=5$ dimensions\footnote{
In the $n=5$ minimal supergravity theory described by~\eqref{CSAction} (and also
in other supergravity theories), it is common to consider solutions possessing a
covariantly constant spinor field.
By forming a suitable bi-linear combination of this spinor field, one
obtains an everywhere non-spacelike Killing vector field, which in particular
must be null at the event horizon. Therefore, the event horizon in minimal supergravity
theories in $5$-dimensions must be a Killing horizon for such solutions~\cite{GGHPR2002,Reall03}.
}.
%(In that case we are already done.) Therefore, in principle, we are
%concerned with {\sl non-supersymmetric} theories described by
%the above action.]
This theory has a metric and a single gauge field with field strength tensor
$F_{ab}= \nabla_aA_b-\nabla_bA_a$; we set the Fermi-fields equal to zero.
Its action is
\ben\label{CSAction}
 S = \int {\dd}^5 x \sqrt{-g}
                             \left(R - F_{ab}F^{ab}
            + \frac{2}{3\sqrt{3}}
              \epsilon^{abcde} F_{ab} F_{cd} A_e  \right) \,.
\een
The last term in this action is a Chern-Simons term.
The resulting Einstein equations (i.e., varying $g_{ab}$)
are precisely the same as those given previously in eqs.~(\ref{eq:grav})
and (\ref{def:Tab}) with $\phi^i=0$ and $h_{IJ}=\delta_{IJ}$,
as the stress-energy tensor is not modified by the addition of
the Chern-Simons term, whereas the equations of motion  for the gauge field (i.e., varying $A_a$)
are modified to
\ben
 \nabla_b F^{ba} + \frac{1}{2\sqrt{3} }
                           \epsilon^{a bcde}
                             F_{bc} F_{de} =0\,.
\label{eom:gauge:CS}
\een
We decompose this equation by contracting the free index into
$n^a$, $\ell^a$, and $p^a{}_b$, respectively.
The first term of the left-hand side of the above equation is given
by eqs.~(\ref{eom:gauge:u}), (\ref{eom:gauge:r}), and (\ref{eom:gauge:C}),
respectively (with $h_{IJ}= \delta_{IJ}$), and the second term is given respectively by
\bena
\ell_a \epsilon^{a bcde} F_{bc} F_{de}
&=&
 - 4 \epsilon_{/\!/}^{abc} W_a U_{bc} \,,
\label{cs:u}
\\
(\dd r)_a \epsilon^{a bcde} F_{bc} F_{de}
&=& 4 \epsilon_{/\!/}^{abc}\left(
                          V_aU_{bc} - r \cdot U_{ab}U_{cd}\beta_e q^{de}
                    \right) \,,
\label{cs:r}
\\
p^a{}_e \epsilon^{ebcdf} F_{bc} F_{de}
&=& - 4 \epsilon_{/\!/}^{abc} \left(SU_{bc} -2 V_b W_c \right)
\non \\
  &+& 4 r \cdot \epsilon_{/\!/}^{ebc}
       \Big\{
              - q^{af}\beta_f  U_{bc} W_e
              +  p^a{}_e q^{df} \beta_f
                 ( U_{bc} W_d + 2 W_b U_{cd} )
       \Big\} \,,
\label{cs:A}
\eena
where we have introduced
$
 \epsilon_{/\!/}^{abc} = q^{ad}q^{be}q^{cf}n^p\ell^q\epsilon_{pqdef}
$.

\medskip
We will now outline how to prove the rigidity theorem in the
presence of the Chern-Simons term. For brevity, we only outline
the main changes compared to the case without Chern-Simons term described in Sect.~\ref{sect:Matterfields}.
Recall that in our proof we need to use the equation of motion for the gauge field only when
we show the following equations:
\bena
\underbrace{\pounds_\ell \, \pounds_\ell \, \cdots \,
\pounds_\ell}_{m \,\, {\rm times}}
\left(\pounds_n S  \right) = 0 \, , \quad
\underbrace{\pounds_\ell \, \pounds_\ell \, \cdots \,
\pounds_\ell}_{m \,\, {\rm times}}
\left(\pounds_n W_a \right) = 0 \,.
\label{eqs:proof:CS}
\eena
% The proof of eqs.~(\ref{cid}), (\ref{cid:s}), and (\ref{cid4}) are
% irrelevant to the existence of the Chern-Simons term.
First we note that since $V_a =0$ on $H$ from the Raychaudhuri equation,
eq.~(\ref{cs:r}) must vanish on $H$.
Thus, when contracted with $(\dd r)_a$, the Chern-Simons term in
eq.~(\ref{eom:gauge:CS}) is irrelevant to the equation of motion at least
on $H$. We can then show that the first of eq.~(\ref{eqs:proof:CS}) is
satisfied for $m=0$, so
$\pounds_n S=0$ on $H$. We can then easily show that the Lie-derivative
$\pounds_n$ of eq.~(\ref{cs:A}) vanishes on $H$ and hence does not contribute to
the Lie-derivative $\pounds_n$ of eq.~(\ref{eom:gauge:CS})
contracted with $p^c{}_b$ on $H$. Then, from these results, we find that
eq.~\eqref{eq:WWV} also holds in the presence of a Chern-Simons term. By the
same argument as after eq.~\eqref{eq:WWV}, we then conclude that
the second of eq.~(\ref{eqs:proof:CS}) holds for $m=0$.
Next, taking the Lie-derivative $\pounds_n$ of
eq.~(\ref{eom:gauge:CS}) contracting it with $\ell_a=(\dd u)_a$
and using the results derived so far (in particular $\pounds_n W^a=0$ on $H$),
we can show that the first of eq.~(\ref{eqs:proof:CS}) holds for
$m=1$. Furthermore, taking $\pounds_n \pounds_\ell$ of eq.~(\ref{eom:gauge:CS})
contracted with $p^c{}_b$, and using what we already know (in particular
$\pounds_n \pounds_\ell S=0$ on $H$), we can show that the second
of eq.~(\ref{eqs:proof:CS}) holds for $m=1$. Then, using
inductive method as in a similar manner for the vacuum case, we can show
that eqs.~(\ref{eqs:proof:CS})---as well as eqs.~(\ref{cid}),
(\ref{cid:s}), and (\ref{cid4})---hold for all $m=0,1,2,\dots$.
Thus, our rigidity theorems also applies to the theory described by the
action~\eqref{CSAction}.

\end{document}